\documentclass[conference,dvipsnames]{IEEEtran}
\IEEEoverridecommandlockouts

\usepackage{cite}
\usepackage{amsmath,amssymb,amsfonts}
\usepackage{algorithm2e}
\usepackage{graphicx}
\usepackage{textcomp}
\usepackage{xcolor}
\usepackage{paralist}
\usepackage{circledsteps}
\usepackage{makecell}
\usepackage{url}
\usepackage{hyperref}
\usepackage[most]{tcolorbox}
\usepackage{cleveref}
\usepackage{booktabs}
\usepackage{amssymb}
\usepackage{colortbl}
\usepackage{array}

\def\BibTeX{{\rm B\kern-.05em{\sc i\kern-.025em b}\kern-.08em
    T\kern-.1667em\lower.7ex\hbox{E}\kern-.125emX}}

\newcommand{\cmark}{$\checkmark$}
\newcommand{\pmark}{\raisebox{0.15ex}{\small$\circ$}}

\newcommand{\rrr}{$\text{R}^3$\xspace}
\newcommand{\rrrfull}{Record-Remix-Replay\xspace}

\definecolor{ObservationBlue}{HTML}{006699}
\newtcolorbox{observationboxnonumber}[1][]{%
  enhanced,
  breakable,
  colback=gray!3,
  colframe=black!20,
  coltitle=black,
  fonttitle=\bfseries,
  title={Observation},
  boxrule=0.5pt,
  leftrule=0pt,
  rightrule=0pt,
  toprule=0pt,
  bottomrule=0pt,
  borderline west={2pt}{0pt}{ObservationBlue},
  boxsep=4pt,
  left=7pt,
  right=7pt,
  top=5pt,
  bottom=5pt,
  #1
}

\newtcolorbox[auto counter]{observationbox}[2][]{%
  enhanced,
  breakable,
  colback=gray!3,
  colframe=black!20,
  coltitle=black,
  colbacktitle=gray!10,
  coltitle=MidnightBlue!50,
  fonttitle=\bfseries,
  title={Observation~\thetcbcounter: #2},
  boxrule=0.5pt,
  leftrule=0pt,
  rightrule=0pt,
  toprule=0pt,
  bottomrule=0pt,
  borderline west={2pt}{0pt}{ObservationBlue},
  boxsep=4pt,
  left=7pt,
  right=7pt,
  top=5pt,
  bottom=5pt,
  #1
}
    
\begin{document}

\title{Record-Remix-Replay: Hierarchical GPU Kernel Optimization using Evolutionary Search}

\author{
\IEEEauthorblockN{
Daniel Nichols\IEEEauthorrefmark{1},
Konstantinos Parasyris,
Caetano Melone,
Tal Ben-Nun,
Giorgis Georgakoudis,
Harshitha Menon
}
\IEEEauthorblockA{
\textit{Lawrence Livermore National Laboratory} \\
Livermore, USA \\
\{danielnichols, parasyris1, cmelone, talbn, georgakoudis1, harshitha\}@llnl.gov
}
\IEEEauthorblockA{\IEEEauthorrefmark{1}Corresponding author}
}

\maketitle

\begin{abstract}
As high-performance computing and AI workloads become increasingly dependent on GPUs, maintaining high performance across rapidly evolving hardware generations has become a major challenge. 
Developers often spend months tuning scientific applications to fully exploit new architectures, navigating a complex optimization space that spans algorithm design, source implementation, compiler flags and pass sequences, and kernel launch parameters. 
Existing approaches can effectively search parts of this space in isolation, such as launch configurations or compiler settings, but optimizing across the full space still requires substantial human expertise and iterative manual effort.
In this paper, we present \emph{\rrrfull (\rrr)}, a hierarchical optimization framework that combines LLM-driven evolutionary search, Bayesian optimization, and record-replay compilation techniques to efficiently explore GPU kernel optimizations from source-level implementation choices down to compiler pass ordering and runtime configuration. 
By making candidate evaluation fast and scalable, our approach enables practical end-to-end search over optimization dimensions that are typically treated separately. 
We show that \emph{\rrrfull} can optimize full scientific applications better than traditional approaches over kernel parameters and compiler flags, while also being nearly an order of magnitude faster than modern evolutionary search approaches.

\end{abstract}

\section{Introduction}\label{sec:intro}

GPUs power much of today's high-performance computing and large-scale AI workloads, and their efficiency increasingly determines the cost, energy consumption, and time-to-solution of scientific and industrial computing. 
Small improvements in GPU utilization can translate into substantial savings at cluster scale and enable higher-fidelity simulations or faster model training.
Yet finding these small improvements is notoriously difficult: modern GPU software stacks expose a deep hierarchy of optimization opportunities ranging from algorithm choices and implementation to compiler optimizations and launch parameters.
Finding ideal kernels within this space is usually an arduous and slow trial-and-error task left to computational scientists and performance engineers.

Despite decades of progress in performance engineering and auto-tuning, GPU optimization remains a critical bottleneck in practice because existing methods address only parts of the problem. 
Domain experts can reason about algorithmic and implementation-level changes, while auto-tuners are effective over structured spaces such as launch parameters, tiling choices, and selected compiler options. 
However, large performance gains often require improvements that span these layers simultaneously. 
As GPU systems become more important to scientific computing and AI, this fragmentation becomes increasingly costly: it slows optimization cycles, limits portability across architectures, and leaves substantial performance untapped in expensive production workloads. 
A method that can systematically optimize across the full vertical stack of GPU kernel decisions would therefore be highly valuable, both for improving application performance and for reducing the manual expertise required to obtain it.

Optimizing across the entire range of GPU kernel design decisions is difficult, however.
Algorithmic structure, kernel implementation, compiler decisions, and launch parameters all shape the final execution behavior, and their effects are strongly interdependent. 
Small source-level changes can alter register pressure, memory access patterns, or synchronization structure, which in turn change which compiler transformations are profitable and which launch configurations are effective. 
This creates a search space that is large, hierarchical, and non-separable: good choices at one layer depend on choices made at the others. 
Consequently, approaches that optimize only launch parameters, only compiler passes, or only source rewrites leave substantial performance opportunities unexplored.

At the same time, exploring this space is expensive because performance feedback is slow to obtain and difficult to stabilize. 
For large GPU applications, evaluating a candidate optimization often requires rerunning a substantial portion of the application, often several times to account for measurement noise.
These long feedback cycles make broad search impractical. 
They are especially problematic for methods that require many evaluations, such as evolutionary optimization, and for methods that generate diverse source variants, such as LLM-based coding agents. 
Classical auto-tuners mitigate this cost by restricting search to structured parameter spaces, but that restriction also limits the kinds of transformations they can discover. 
What is missing is a way to make evaluation cheap enough to support rapid search while still covering the full space of kernel optimizations.

Our paper introduces \rrrfull (\rrr) that combines evolutionary optimization using large language models (LLMs) with Bayesian optimization, enabling us to search optimal source code, launch configurations, and compiler passes.
To overcome the massive runtime requirement for evaluating applications thousands of times, we utilize record-replay techniques alongside several novel optimizations to accelerate and scale GPU kernel optimization.
We find that \rrr is able to accelerate LLM-based evolutionary search by nearly $10\times$ and tune GPU kernels to better speedups than existing approaches.
Using \rrr we are able to reduce the time spent in compute in a large, production HPC application by over 28\%.

Our paper makes the following important contributions:

\begin{itemize}
    \item We introduce a novel approach combining LLM evolutionary search and record-replay to rapidly explore an optimization space including launch parameters, compiler passes, and implementations/algorithms.
    \item We introduce novel improvements to existing record-replay techniques to enable rapid replay of general CUDA and HIP kernels.
    \item We introduce several novel optimizations to enable scaling evolutionary optimization to hundreds of GPUs.
    \item We release our record-replay contributions in the open-source framework Mneme (\href{https://github.com/Olympus-HPC/Mneme}{https://github.com/Olympus-HPC/Mneme}). \rrr is planned to be open-sourced in the coming weeks.
    \item We conduct evaluations of our approach on several multi-kernel scientific applications. \rrr is able to find a 28.4\% reduction in compute time in the QUDA scientific code.
\end{itemize}

\section{Background}\label{sec:bg}
This section presents background on auto-tuning, record-replay compiler instrumentation techniques, and evolutionary optimization powered by LLMs.

\subsection{Auto-tuning}\label{sec:bg:auto_tuning}
Auto-tuning refers to techniques that automatically search for program configurations to optimize a given performance objective, such as execution time, energy consumption, or resource utilization.
Rather than relying solely on manual performance engineering, auto-tuners explore a space of possible implementations or parameter settings that influence program behavior. 
These parameters may include algorithmic variants, compiler optimization flags, tiling factors, memory layouts, or runtime launch parameters. 
Most auto-tuning systems operate iteratively: in each iteration, the tuner generates a candidate configuration, evaluates it by executing the program (or a representative kernel), and uses the measured performance as feedback to guide subsequent choices. 
The decision process may be driven by strategies such as random search, evolutionary algorithms, or model-based optimization methods. 
Through repeated evaluation and feedback, auto-tuning progressively identifies configurations that improve the target performance metric.

Many auto-tuning frameworks have been proposed for GPU and HPC applications, including general-purpose systems such as OpenTuner~\cite{7855909}, ActiveHarmony~\cite{1592880}, Kernel Launcher~\cite{heldens2023kernel}, and CLTune~\cite{7328205}. 
These frameworks differ in the parameters they optimize and the search strategies they employ, but all rely on iterative evaluation of candidate configurations. 
Because each candidate typically requires executing the application/kernel, the evaluation cost often becomes a dominant limitation.

\subsection{Record-Replay}\label{sec:bg:record_replay}

Record-and-replay decouples optimization \emph{evaluation} from full-application execution by recording a \emph{replay unit} together with the \emph{recorded execution context} needed to reproduce it, then replaying transformed variants in isolation.
A replay-based evaluation loop therefore consists of: (1) selecting a replay unit, (2) capturing sufficient state to reconstruct its execution, and (3) re-executing that unit repeatedly under modified code, compiler settings, or runtime parameters.
Its key benefit is lower evaluation cost: replay avoids rerunning unrelated application code, enables selective tuning of hotspots, and exposes parallelism across replay instances.

Prior work instantiated this model at different granularities.
CERE~\cite{10.1007/978-3-319-43659-3_18,DBLP:journals/taco/CastroAPPJ15,10.1145/3572848.3577497} uses CPU \emph{codelets} as replay units, extracting hotspot regions at LLVM IR level so they can be modified, recompiled, and replayed independently from the original program.
To keep replay representative while reducing cost, CERE captures the working set and cache state and selects representative invocations.
In contrast, Parasyris et al.~\cite{parasyris_scalable_20232} use the \emph{GPU kernel invocation} itself as the replay unit.
Their recorded execution context includes kernel launch arguments, launch configuration, device memory state, relevant globals, and the kernel image; replay reconstructs this state and executes the kernel as a standalone artifact.
Compared to codelet replay, kernel replay is a more natural granularity for GPU tuning because it aligns directly with the unit whose compiler transformations and launch parameters determine performance.
Because replay instances are self-contained, they can be tuned selectively, migrated to compatible systems, and evaluated in parallel, substantially reducing optimization cost.

Our work builds on this replay-based evaluation model, using GPU-kernel replay as the inner evaluation engine for hierarchical search across source implementations, compiler transformations, and launch configurations.

\subsection{Evolutionary Optimization with LLMs}\label{sec:bg:evolve}
Evolutionary optimization has been a popular optimization technique for many decades.
Consider the task of finding some $x^\star$ that minimizes $f(x)$.
Evolutionary optimization creates populations of candidate $x$ values, evaluates each candidate using $f$ or a proxy fitness function, and then \emph{evolves} the population to find the next candidate $x$ values.
How populations are evolved can take many forms, but is most often based on genetic evolutionary algorithms, where {\it selection}, {\it mutation}, and/or {\it crossover} are used to remove bad candidates, explore changing good candidates, and migrate candidates between populations, respectively.
This style of optimization has been very successful at exploring diverse solutions to optimization problems, particularly non-differentiable ones.
We refer the reader to~\cite{liu_survey_20232} for a more complete discussion and background on evolutionary algorithms.

Up until recently, evolutionary algorithms were only useful for search spaces where mutation was clearly defined, that is, where it is clear how to mutate the current population into the next generation.
For example, discrete and continuous spaces are generally trivial to mutate; we can increment numbers, negate them, flip bits, choose random subsets of discrete sets, etc.
But other search spaces that are expressed as structured text, like algorithms and source code, are much harder to mutate.
We could alter, remove, and/or add new characters, but most mutations would lead to invalid code and the evolutionary search would never converge.
However, the recent increase in language modeling capabilities via LLMs has made it possible to use evolutionary search over a text space.
Recent works like FunSearch~\cite{romera-paredes_mathematical_20242}, AlphaEvolve~\cite{novikov_alphaevolve_20252}, and ADRS~\cite{cheng_barbarians_20252} have shown the impressive capabilities of these systems in discovering new algorithms and systems optimizations.

\begin{figure}
    \centering
    \includegraphics[width=0.9\linewidth]{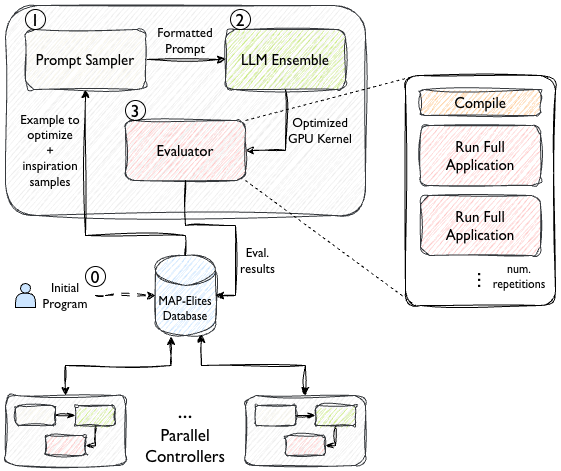}
    \caption{Overview of MAP-Elites evolution as in AlphaEvolve/OpenEvolve. \Circled{1} Prompts are constructed from the MAP-Elites population database, \Circled{2} sent to an LLM randomly selected from an ensemble of LLMs, and \Circled{3} evaluated by compiling and running the code. The population database is then updated according to the MAP-Elites algorithm. Multiple parallel controllers execute this loop concurrently to increase throughput.\label{fig:evolve-diagram}}
\end{figure}

Many of these works, like AlphaEvolve~\cite{novikov_alphaevolve_20252} and its popular open-source reproduction OpenEvolve~\cite{openevolve}, are based on the {\bf MAP-Elites evolutionary algorithm}~\cite{mouret_illuminating_20152}.
\Cref{fig:evolve-diagram} presents an overview of the MAP-Elites evolutionary algorithm powered by LLMs for code optimization.
In the MAP-Elites algorithm, members of the population are mapped to coordinate cells in a cartesian grid based on ``feature dimensions", i.e. each dimension of the grid corresponds to a feature of solutions in the search space.
Common feature dimensions are complexity (e.g. lines of code) and diversity (e.g. hamming distance from other samples).
When a new solution is sampled, it is evaluated and mapped to a coordinate cell based on its features.
If there is already a solution in that cell and it is worse than the new one, we replace it.
The solutions in each coordinate cell are called ``elites".
We keep mutating existing samples, evaluating the mutations, and adding them back into the grid until we reach a desirable convergence criterion.
MAP-Elites was first introduced to increase the diversity of samples explored during evolutionary optimization.

Each island maintains its own MAP-Elites database in which multiple {\it parallel controllers} concurrently execute the sample, evaluate, and update cycle.
When multiple islands are used we employ {\it migration} to move data between them; migration occurs regularly, e.g. every $N$ iterations, and moves samples between MAP-Elites databases.
One popular migration implementation is to, whenever migration occurs, replace the bottom half of samples in every island, with one of the top samples from the other islands.

For GPU kernel optimization, the main limitation of this approach is evaluation cost.
Each iteration still requires compiling and running the candidate to measure correctness and performance.
For large GPU applications, where a single evaluation may require substantial hardware and long execution times, this makes LLM-guided MAP-Elites prohibitively expensive.
This evaluation bottleneck is precisely what motivates our use of record-replay as the inner evaluation engine in \rrr.

\section{Related Work}\label{sec:related-work}

In this section we present related work on auto-tuning, record-replay, and evolutionary based GPU kernel optimization methods.
We focus solely on works used for GPU kernel optimization as that is the target of our work.

\newcommand{\tablesection}[1]{%
\addlinespace[5pt]
\rowcolor{black!6}
\multicolumn{9}{@{}l}{\color{black!80}\textsc{#1}}\\[-2pt]
\addlinespace[2pt]
}

\begin{table*}[t]
\centering
\small
\setlength{\tabcolsep}{3.0pt}
\renewcommand{\arraystretch}{1.08}
\caption{
Representative related work in GPU kernel auto-tuning and optimization.
\cmark\ = first-class support; \pmark\ = technically possible but indirect / not the primary design point.
Evaluation cost is qualitative and reflects the typical inner-loop tuning cost:
\emph{Low} = isolated kernel / replay / JIT evaluation,
\emph{Med.} = operator or graph-level tuning,
\emph{High} = full-application.
}
\label{tab:related_work_matrix}

\begin{tabular}{@{}p{6.04cm}ccccccc>{\centering\arraybackslash}p{2.0cm}>{\centering\arraybackslash}p{0.95cm}@{}}
\toprule
& \multicolumn{2}{c}{\textbf{Source-level}} &
\multicolumn{2}{c}{\textbf{Compiler-level}} &
\textbf{Runtime} &
\textbf{System} &
\multicolumn{2}{c}{\textbf{Eval.}}  \\
\cmidrule(lr){2-3}\cmidrule(lr){4-5}\cmidrule(lr){6-6}\cmidrule(lr){7-7}\cmidrule(lr){8-9}
\textbf{Work} &
\makecell[c]{Unstruct.\\rewrite} &
\makecell[c]{DSL /\\template} &
\makecell[c]{Flags /\\params} &
\makecell[c]{Pass\\order} &
\makecell[c]{Launch\\cfg.} &
\makecell[c]{Env. /\\resources} &
\makecell[c]{Required \\Execution} &
\makecell[c]{Cost} 
\\
\midrule

\tablesection{Kernel parameter auto-tuners}
CLTune; Kernel Tuner~\cite{7328205,van2019kernel} &  & \cmark &  &  & \cmark & \pmark & Kernel & Low \\
Kernel Launcher~\cite{heldens2023kernel} &  & \cmark &  &  & \cmark &  & Kernel & Low \\

\tablesection{DSL auto-schedulers}
TVM / Ansor / MetaSchedule~\cite{chen_tvm_20182,zheng_ansor_2020,shao_tensorprob_2022} &  & \cmark & \pmark &  & \cmark &  & Kernel / Graph & Med. \\
Halide autoschedulers~\cite{adams_halide_autoscheduler_2019} &  & \cmark &  &  & \cmark &  & Kernel & Low \\
Triton~\cite{tillet_triton_2019} &  & \cmark & \pmark & \pmark & \cmark &  & Kernel & Low \\

\tablesection{Compiler auto-tuners}
OpenTuner (e.g., GCC flags)~\cite{7855909} &  &  & \cmark &  &  & \pmark & App. & High \\
MiCOMP; CompilerGym~\cite{10.1145/3124452,cummins_compilergym_2022} &  &  &  & \cmark &  &  & App. & High \\

\tablesection{Runtime/system tuners}
ActiveHarmony; Apollo; Artemis~\cite{1592880,beckingsale2017apollo,10.1007/978-3-030-78713-4_24} &  &  &  &  & \pmark & \cmark & App. & High \\
HiPerBOt; GPTune; Periscope~\cite{menon_hiperbot_2020,10.1145/3437801.3441621,mijakovic_ptf2_2016} &  &  & \cmark &  &  & \cmark & App. & High \\

\tablesection{Replay / JIT substrates}
CERE (codelets)~\cite{DBLP:journals/taco/CastroAPPJ15} &  & \pmark & \cmark & \pmark &  & \cmark & Replay Unit & Low \\
Kernel RR~\cite{parasyris_scalable_20232} & \pmark &  &  &  & \cmark &  & Replay Unit & Low \\
Proteus (LLVM-IR JIT)~\cite{proteus} &  &  &  & \pmark & \cmark &  & Kernel & Low \\

\tablesection{LLM evolutionary agents}
FunSearch; AlphaEvolve~\cite{romera-paredes_mathematical_20242,novikov_alphaevolve_20252} & \cmark &  & \pmark & \pmark & \pmark &  & App. & High \\
\midrule

\rowcolor{gray!10}
\textbf{\rrrfull (this work)} &
\cmark &  & \cmark & \cmark & \cmark &  & \textbf{Replay Unit} & \textbf{Low} \\
\bottomrule
\end{tabular}
\end{table*}

Prior work in GPU kernel optimization typically targets only one layer of the tuning stack (\Cref{tab:related_work_matrix}). Kernel parameter auto-tuners (e.g. CLTune) focus on structured template choices and launch configurations, and therefore achieve low-cost kernel-scope evaluation~\cite{7328205,van2019kernel,heldens2023kernel}. DSL auto-schedulers (e.g. TVM) broaden the structured search space, but require adoption of a DSL or specialized IR and still primarily optimize at kernel or graph scope~\cite{chen_tvm_20182,zheng_ansor_2020,shao_tensorprob_2022,adams_halide_autoscheduler_2019,tillet_triton_2019}. In contrast, compiler auto-tuners (e.g. CompilerGym) search compiler flags or phase orderings, while runtime and system tuners (e.g. ActiveHarmony) tune execution or platform parameters; these methods generally operate at application scope and incur higher evaluation cost~\cite{7855909,10.1145/3124452,cummins_compilergym_2022,1592880,beckingsale2017apollo,10.1007/978-3-030-78713-4_24,menon_hiperbot_2020,10.1145/3437801.3441621,mijakovic_ptf2_2016}.

Replay and JIT substrates (e.g. CERE) reduce evaluation cost substantially by isolating codelets or kernels as replay units, but they do not search unstructured source-level transformations~\cite{DBLP:journals/taco/CastroAPPJ15,parasyris_scalable_20232,proteus}. 
Conversely, LLM-based evolutionary systems (e.g. AlphaEvolve) or kernel optimization agents (e.g. opt-r1) can explore unstructured rewrites, but typically rely on full-application compilation and execution and therefore remain expensive for large GPU applications~\cite{romera-paredes_mathematical_20242,novikov_alphaevolve_20252,openevolve,nichols2025integratingperformancetoolsmodel,rlpf}. 
\rrr combines the strengths of these lines of work: it uses an outer LLM/MAP-Elites loop to search unstructured kernel rewrites, while an inner record-replay loop provides low-cost evaluation and Bayesian optimization over compiler and launch parameters. 
Quantitatively, our closest baselines are OpenEvolve and Kernel RR-style BO.

\section{\rrr: \rrrfull Overview}\label{sec:r3-overview}
\begin{figure}[h]
    \centering
    \includegraphics[width=0.98\linewidth]{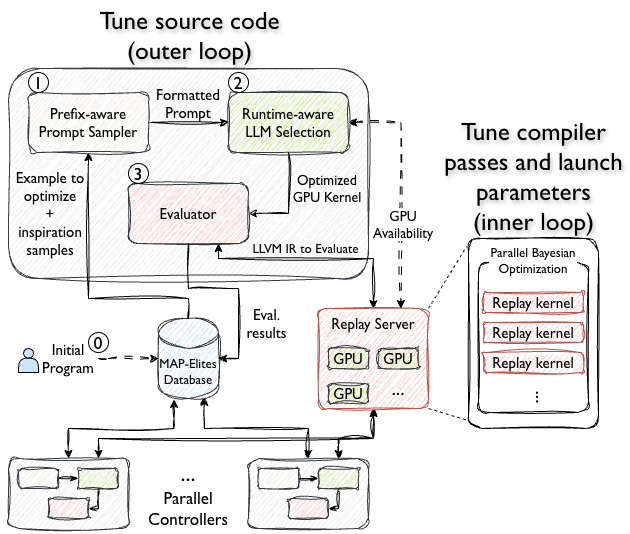}
    \caption{Overview of our \rrrfull framework. \Circled{1} First, the prefix-aware prompt sampler is used to construct prompts that optimize prefix cache hits. \Circled{2} An LLM is selected intelligently based on estimated generation time to maximize GPU utilization and is used to optimize the GPU kernel. \Circled{3} LLVM IR is generated for the optimized kernel and sent to the replay server, which replays kernels and runs parallel Bayesian optimization to find the best runtime across compiler passes and launch parameters. The population database is then updated according to the MAP-Elites algorithm.
    \label{fig:r3-diagram}}
\end{figure}

In this section we detail our proposed \rrrfull (\rrr) framework.
The framework combines LLM-powered MAP-Elites evolution to optimize the source code implementation of GPU kernels with record-replay and Bayesian optimization (BO) to tune the launch configuration and compiler passes for kernel implementations.
\Cref{fig:r3-diagram} provides an overview of the \rrr framework.

\rrr decomposes kernel tuning into two levels to most naturally fit the hierarchy of kernel tuning decisions: text-based source level transformations and structured numeric search over compiler passes and launch configurations.
At the outer level, we use MAP-Elites evolution to search over source-level kernel implementations. 
At the inner level, for each source candidate, we use BO + replay-based auto-tuning to search over lower-level decisions such as compiler optimization pipelines and launch parameters. 
This structure allows the framework to explore algorithmic and implementation changes without giving up the efficiency of structured auto-tuning where it is most effective.
Bayesian optimization has previously demonstrated strong success in searching over structured kernel hyperparameters~\cite{parasyris_scalable_20232}, but is not well suited towards searching new source code implementations.
Similarly, MAP-Elites evolution {\it could} be used to guide LLMs to generate launch configurations, but we experimentally find that standard BO approaches are much more efficient and effective for this task than LLMs.

The input to \rrr is an initial implementation of the target kernel together with a set of recorded executions obtained from representative application runs (see~\Cref{sec:bg:record_replay} for record-replay context).
This implementation is what we will be optimizing and becomes the initial MAP-Elites population (step~\Circled{0}).
At each iteration of the algorithm, the current MAP-Elites population is used to construct prompts that will be used to prompt an LLM to generate optimizations (step~\Circled{1}).
These prompts will contain a candidate kernel implementation that needs to be optimized, historical good and bad examples from the database, and necessary context about the current task, so that the LLM knows to output an optimized kernel.
Rather than forming prompts arbitrarily, \rrr uses a prefix-aware prompt sampler that increases reuse of static prompt prefixes across generations. 
This reduces generation overhead and improves throughput when many candidate mutations are produced over the course of evolution.

Given a formatted prompt, the framework selects an LLM from an ensemble of available models and uses the selected model to propose a new kernel implementation (step~\Circled{2}).
The model choice is runtime-aware: rather than assigning requests uniformly, \rrr schedules generations based on estimated latency and current device availability so that inference resources remain well utilized. 
Once a model is selected it is used to generate an optimized version of the kernel in its prompt.

The generated candidate kernel is then evaluated to determine its correctness and efficiency (step~\Circled{3}).
Evaluation begins by compiling the generated candidate to LLVM IR, which serves as the interface between source-level evolution and replay-based evaluation.
We note that this itself is a significant optimization over standard evolution where re-compilation can be a huge bottleneck, particularly if any changed code is in a header that many translation units depend on or if there are many files that need to be linked. %
By getting the LLVM IR for only the kernel of interest, we can drastically reduce compile and linking time during evaluation.

The compiled LLVM IR for the candidate is then sent to the replay server. 
The replay server reconstructs a previously recorded kernel execution, lowers the candidate IR to an executable kernel, and evaluates it in isolation from the full application (\Cref{sec:mneme-overview} details the record-replay engine).
That is, we ``replay" the GPU kernel as a standalone binary using the same GPU state as in the full application execution.
In \rrr we replay kernels $K$ times (e.g. $10\times$) to obtain statistically significant performance results.
When replaying/evaluating a kernel implementation candidate, the server runs a parallel Bayesian optimization (BO) search over compiler pass configurations and launch parameters to identify the best-performing realization of that implementation. 
We discard any compilation pipelines or launch configurations that lead to the kernel being incorrect.
By utilizing BO, source mutations are not judged by a single arbitrary compiler setting or launch configuration, but by the best configuration that can be found for that candidate.
The replay server returns the best found configuration, its runtime, and the correctness of the kernel (\Cref{sec:mneme:correctness} has more details on how \rrr checks correctness).

Finally, the measured performance and correctness results are returned to the MAP-Elites controller and used to update the population database. 
The resulting elites therefore represent source implementations paired with their correctness and near best possible performance after tuning.
This optimization is run in a loop, repeating \Circled{1},~\Circled{2}, and~\Circled{3} until some finishing criteria is reached.
\rrr supports three different stopping criteria: maximum number of iterations, maximum wall clock time, or after a set number of iterations with no improvement.
The above optimization loop runs for a single GPU kernel and many instances of it are run in parallel for a code with multiple GPU kernels.
In this sense, \rrr is embarrassingly parallel as we can optimize each GPU kernel in parallel with no synchronization or communication.

\section{Record-Replay Engine}\label{sec:mneme-overview}
\rrr utilizes record-replay to rapidly increase the rate at which candidate GPU kernels can be evaluated.
While there are previous record-replay systems, they lack important capabilities needed for use in tight evolutionary optimization.
Relative to prior replay systems, our engine contributes four capabilities that are critical for our setting:
\begin{inparaenum}
\item a persistent replay server that amortizes replay initialization across many evaluations,
\item direct, in-memory operation on recorded LLVM IR through Proteus and LLVM APIs, enabling low-cost compiler and runtime transformations, and
\item correctness checking that is conservative by default but configurable when numerical equivalence rather than bitwise equality is desired.
\item  the ability to record and replay arbitrary CUDA and HIP kernels.
\end{inparaenum}
Together, these make replay practical for powering evaluation for large-scale evolutionary optimization.

The engine follows the standard replay workflow of instrumentation, recording, and replay, but specializes each stage for repeated kernel evaluation.
Instrumentation prepares the application so GPU kernels can be intercepted and preserved as replay units.
Recording selects as \emph{Replay Units} the execution context of representative kernel invocations.
Replay reconstructs that context, applies transformations, and evaluates candidate implementations under alternative compiler and launch settings.

\subsection{Instrumentation and Recording}

Instrumentation is implemented by extending the Proteus JIT infrastructure~\cite{proteus} to identify GPU kernels, extract each kernel together with its transitive dependency closure as LLVM IR, and embed the resulting IR in the application binary.
Kernel launches are redirected through the Proteus runtime so they can later be intercepted and replayed.
We preserve kernels as LLVM IR because it is the representation that both minimizes re-evaluation cost (by reducing compile time) and enables transformation during replay (via LLVM passes).

Recording is performed during representative application runs through a preload library that intercepts kernel launches and device memory operations.
For each recorded kernel invocation, the engine stores: 
\begin{inparaenum}
\item the kernel LLVM IR module, 
\item launch configuration, 
\item kernel arguments, 
\item the pre-execution device state, and 
\item the post-execution device state.
\end{inparaenum}
Using the terminology of~\cref{sec:bg:record_replay}, the pre-execution state forms the \emph{prologue snapshot} used to reconstruct the replay context, while the post-execution state forms the \emph{epilogue snapshot} used for correctness checking.
Because the replay unit is captured independently from the full application, later experiments can be run without rebuilding or re-executing the original code.

\subsection{Replay, Transformation, and Auto-Tuning}

During replay, the engine restores the recorded execution context, recompiles the recorded LLVM IR, and executes the replay unit either with the recorded launch configuration or with alternative settings sampled by the optimizer.
Replay operates directly in memory on LLVM IR through Proteus rather than through a file-based recompilation workflow like in other works~\cite{parasyris_scalable_20232}.
This enables low-cost exploration of transformations such as argument specialization by constant propagation, specialization of thread and block dimensions, insertion of launch bounds, standard LLVM optimization levels, and custom compiler pipelines.
The transformed kernel is then lowered to a device binary and executed on the reconstructed state.

Built on top of replay, the engine exposes a programmatic auto-tuning interface over compiler and runtime parameters.
A sampled point in the search space corresponds to a replay experiment: compile the candidate with a selected transformation pipeline, execute it under a selected launch configuration, and return its measured runtime if correct.
In \rrr this interface is used with Bayesian optimization inside the replay server, but the mechanism itself is agnostic to the search strategy.

\subsection{Correctness Checking}\label{sec:mneme:correctness}

Correctness is defined relative to the recorded execution context.
Given a replay instance, the engine restores the prologue snapshot, executes the candidate kernel, and compares the resulting state against the recorded epilogue snapshot.
By default, this comparison is bitwise exact: the GPU memory state after replaying the kernel must match the memory state from the full program run \emph{exactly}.
This conservative policy ensures that accepted transformations preserve the observed kernel behavior on the recorded state.

Some optimizations, however, are numerically valid while not bitwise identical, especially for floating-point code.
A common example we encounter in this work is replacing an expression such as \texttt{a / sqrtf(b)} with \texttt{a * rsqrtf(b)}. 
To support such cases, the engine also provides an optional relaxed mode in which selected output variables may be checked with user-provided numerical predicates, such as absolute or relative tolerances.
Such memory addresses are marked in source code with an \texttt{annotate} function.
We emphasize that bitwise correctness checking is the default in our engine; checking correctness with relaxed error tolerances requires users to explicitly annotate what variables can have relaxed constraints and what the strictness of those constraints are.
Candidates that fail to compile, crash, or violate the correctness criterion are classified as incorrect and rejected.

To balance speed and robustness, correctness is checked at two granularities.
During search, candidates are typically validated against a single recorded prologue/epilogue pair to keep the inner loop fast.
Promising final candidates are then validated against additional recorded instances of the same kernel when available.
This preserves low evaluation cost during search while ensuring the candidate kernel is correct for all available recorded code paths. %

\subsection{Persistent Replay Server}

A major systems contribution and optimization within our replay engine is a persistent replay server designed for repeated evaluation.
Prior replay systems are effective for offline tuning, but repeated kernel evaluation still suffers from overheads such as re-initializing device memory, reloading replay state, materializing intermediate files, and transferring correctness checks back to the host.
These costs are a major bottleneck when in the inner loop of evolutionary search.

To address this, \rrr uses a replay server that keeps replay state resident across many evaluations and serves replay requests over a lightweight interface.
On startup, the server loads a database of recorded kernels, maps replay units across available GPU resources, and pre-allocates prologue and epilogue buffers to amortize allocation and data movement costs.
The server then exposes replay and tuning endpoints that accept candidate LLVM IR, reconstruct the appropriate replay context, execute the requested experiment, and return runtime and correctness results.
Because replay workers persist across requests, initialization overheads are paid once rather than once per candidate.

This design is necessary for replay to be suitable for \rrr.
It increases throughput enough that candidate generation, rather than replay, becomes the dominant bottleneck in many settings, which in turn motivates the candidate generation optimizations introduced in~\Cref{sec:r3-scaling}.

\subsection{Deployment}

After tuning, the engine exports the best-performing configuration for each kernel as a JSON specification describing the selected transformations, compiler settings, and launch parameters.
This specification is consumable by the Proteus runtime, which applies the prescribed optimizations when the original application is executed.
As a result, tuning can be performed offline on representative replay instances, while deployment remains transparent to the application source code.

\section{Scaling Evolutionary Optimization}\label{sec:r3-scaling}
The use of record-replay and our many optimizations in \rrr rapidly increases the rate of evaluation (see~\Cref{sec:mneme-overview}) and makes evolution inference-bound.
Thus, in this section we present optimizations in \rrr that accelerate inference via prompting strategies and runtime-aware inference scheduling.

\subsection{Prefix-cache Aware Prompting}\label{sec:prefix-caching}

Most modern LLM inference engines employ {\it prefix caching}, which reuses intermediate computations when successive generations share a common prompt prefix.
Prefix caching is most beneficial when the {\it prefill} phase is non-negligible compared to {\it decode} time, i.e. for large input prompts and small/moderate outputs.
This matches our setting well: evolutionary prompts contain several large code samples from the database, while the model only needs to generate a diff for a single GPU kernel.
However, we find that prompting strategies used in open-source AlphaEvolve-style frameworks are not prefix-cache friendly and place dynamic information early in the prompt, leading to frequent cache misses.

To address this, we propose a prefix-cache aware prompt sampling algorithm for MAP-Elites-style evolution.
First, we place all static prompt content at the beginning and move dynamic content, such as sampled code examples, to the end.
This ensures that the static portion of the prompt can be reused across generations.

Second, we reorder sampled inspiration codes to maximize reuse of previously cached prompt prefixes.
Given a sampled set of inspiration codes $E = [e_1, \dots, e_k]$, let $\mathcal{H}$ denote the set of recent prompt orderings.
For each $H \in \mathcal{H}$, we find the longest prefix $H[1:\ell]$ that is an ordered subset of $E$ and select the ordering with the largest such $\ell$.
If a match is found, we place that shared prefix first and append the remaining sampled codes in their original sampled order; otherwise, we keep $E$ unchanged.
The final prompt is then formed as a fixed static prefix followed by this reordered list of inspirations.
Because this procedure changes only the order in which sampled codes are presented, and not which codes are sampled, it preserves the underlying MAP-Elites convergence behavior while increasing the likelihood of prefix-cache hits.

\subsection{Runtime Aware Scheduling}\label{sec:prefix-caching}

After record-replay reduces candidate evaluation time, the main bottleneck in \rrr shifts from execution to generation.
A naive way to increase throughput is to issue more parallel LLM requests, but this also increases {\it staleness}: prompts are constructed from a population snapshot while many generations and evaluations are still in flight.
Some staleness is tolerable, but too much reduces sample quality and wastes inference and evaluation resources on candidates proposed from already superseded elites.

A straightforward way to control staleness is to increase the number of islands together with the number of parallel controllers.
This mitigates contention on any single database, but it is not ideal.
With too many islands, each island receives fewer updates and will require more iterations to converge.
Previous works~\cite{romera-paredes_mathematical_20242, novikov_alphaevolve_20252} find $\approx 5$ islands to be ideal and frameworks like OpenEvolve~\cite{openevolve} set recommended defaults between 3 and 5 islands.
In practice, we would like to keep the number of islands relatively small for good convergence behavior while still scaling candidate throughput across many inference and evaluation devices.

To do this, \rrr uses a runtime-aware scheduling policy for heterogeneous LLMs.
We alter the LLM sampling method (step \Circled{2} in~\Cref{fig:evolve-diagram}) to sample LLMs based on expected generation latency, the current set of in-flight requests, and expected evaluation availability.
Intuitively, the scheduler prefers to issue a slower request only when doing so will not leave evaluation resources idle, and otherwise fills short gaps with faster models that can complete generation and replay before the slower request returns.
For example, if a slow model (e.g. gpt-5) is generating a new optimization and the GPU it is scheduled to evaluate on is currently sitting idle, then we can try to {\it backfill} this bubble by running a quick model (e.g. gpt-5-mini) and evaluate its output before the slow model is done generating.
This is similar to backfilling in HPC schedulers: short jobs are used opportunistically to improve utilization without delaying longer jobs.

Concretely, let $\tau_m$ denote the estimated generation time for model $m$, and let $\hat{t}_{\mathrm{eval}}$ denote the current estimate of replay plus autotuning time for a candidate.
When a controller is ready to issue a new generation request, the LLM sampler examines the outstanding generations and the expected times at which replay workers will become available.
If a fast model can complete generation and have its candidate evaluated within slack time that would otherwise be lost while waiting for a slower model, the fast model is selected.
Otherwise, the scheduler may issue a slower, higher-quality model request.
This policy increases overlap between inference and replay, keeps GPUs busier, and improves end-to-end throughput.

An important design point of this policy is that it increases utilization given a fixed number of parallel controllers and GPUs available.
We do not need to increase the number of parallel controllers and, thus, can improve both inference and replay devices utilization while limiting the number of in-flight samples drawn from stale database state.
As we show in Section~\ref{sec:evolve-performance-results}, once replay has made evaluation cheap, this scheduling layer becomes necessary to continue scaling evolution.
We also highlight the simplicity of this approach: it can be implemented directly on top of OpenEvolve or AlphaEvolve without major evaluation infrastructure changes.
Only the LLM selection mechanism needs to be replaced, which can be trivially done in Python without altering the existing frameworks via monkey patching.

\section{Experimental Setup}\label{sec:exp-setup}
In this section we detail the applications used for evaluation, baselines for comparison, and the settings used to run \rrr.

\subsection{Computing Environment}
All AMD experiments are executed on a cluster with MI300A GPUs using the ROCm 7.0.2 software stack.
NVIDIA experiments are executed on a cluster with H100 GPUs and CUDA 12.
When gpt-oss is locally hosted for an experiment, we utilize vLLM~\cite{kwon_efficient_202322} to serve the model.

\subsection{Applications}

We run experiments across four different scientific applications to evaluate different approaches on how well they can optimize GPU kernels.
Applications are selected to cover a variety of common HPC and scientific computing domains.
We compare optimization results on Lulesh~\cite{lulesh}, miniFE~\cite{lin_assessing_2015}, S3D~\cite{shoc_s3d}, and miniWeather~\cite{miniweather} as they all have CUDA and HIP implementations available as well as many GPU kernels in them.
Implementations are taken from HeCBench~\cite{jin_benchmark_20232}, which has full CUDA and HIP implementations of each application.

\noindent\textbf{LULESH} is a shock hydrodynamics proxy application for explicit Lagrangian methods on unstructured hexahedral meshes. It has 15 GPU kernels. %

\noindent\textbf{MiniFE} is an implicit finite-element proxy application that models matrix generation, assembly, and iterative solution for sparse linear systems. It has 7 GPU kernels. %

\noindent\textbf{S3D} is a turbulent combustion DNS code with detailed chemistry and molecular transport. It has 54 GPU kernels. %

\noindent\textbf{miniWeather} is a weather mini-app for dry compressible non-hydrostatic flow. It has 9 GPU kernels, but only 7 on the data path we evaluate. %

\subsection{Baselines and \rrr Configuration}
We compare \rrr against two baselines: kernel parameter tuning with Bayesian optimization (BO) based on~\cite{parasyris_scalable_20232} and OpenEvolve~\cite{openevolve}.
We select these as they represent state-of-the-art in kernel parameter tuning (passes, launch parameters) and implementation.

\vspace{0.07em}
\noindent\textbf{Record-Replay + BO.}
Many works have previously used variations of BO to optimize GPU kernel launch parameters.
We compare with~\cite{parasyris_scalable_20232}, a recent state-of-the-art work that combines BO with record-replay to tune launch parameters and compiler configurations rapidly.
Since~\cite{parasyris_scalable_20232} can only handle OpenMP-offload kernels and not arbitrary GPU code, we re-implement the same BO search in our record-replay engine.
We search over the same configuration (launch and compiler parameters) for 200 iterations and keep the best runtime at the end.
Runtimes are obtained by running each kernel seven times: the first two as a warmup and the latter five for recording.
We note that, while we recreate the search and replay functionality of~\cite{parasyris_scalable_20232}, our implementation will actually be \emph{faster} due to its many optimizations (see~\Cref{sec:mneme-overview}).

\vspace{0.07em}
\noindent\textbf{OpenEvolve.}
To compare with recent LLM-powered evolutionary optimization works we run OpenEvolve~\cite{openevolve} (an open-source reproduction of the closed-source AlphaEvolve~\cite{novikov_alphaevolve_20252}).
OpenEvolve implements a MAP-Elites based evolutionary search process as shown in~\Cref{fig:evolve-diagram}.
We run OpenEvolve for 200 iterations with four islands, a population size of 20, and migrations every 20 iterations; these settings are based on findings in~\cite{romera-paredes_mathematical_20242, novikov_alphaevolve_20252}.

\vspace{0.07em}
\noindent\textbf{\rrrfull (\rrr).}
We run \rrr with the same MAP-Elites settings as OpenEvolve: 200 iterations, four islands, population of 20, and migrations every 20 iterations.
Within the replay server we tune kernels across compiler pass order and launch parameters using parallel Tree Parzen Estimation~\cite{bergstra_algorithms_20112} for 30 iterations as the BO search implementation.
We utilize the \texttt{annotate} feature in \rrr (see~\Cref{sec:mneme:correctness}) to implement correctness constraints for each application (we have further manually verified the correctness of all final optimized kernels by checking they yield the expected scientific result in the full application).

For both OpenEvolve and \rrr we utilize an LLM ensemble of gpt-oss-120b~\cite{openai_gpt-oss-120b_20252}, gpt-5-mini, and gpt-5~\cite{singh_openai_20252}.
OpenEvolve randomly samples these with probabilities 0.6, 0.3, and 0.1, respectively.
We choose these values based on the 90-10 split in~\cite{novikov_alphaevolve_20252} where a large, frontier LLM is used 10\% of the time (Gemini-Pro) and a small, fast LLM is used 90\% of the time (Gemini-flash); this mixture blends fast iteration with high quality optimizations while keeping costs low.
We further split the cheap, fast LLM category into a commercial model and local model to increase output diversity and minimize API costs.
\rrr begins with the same sampling probabilities for LLM selection, but then implements runtime-aware selection on top.
To make fair time-to-solution comparisons we run all methods (record-replay + BO, OpenEvolve, and \rrr) with $4\cdot K$ GPUs, where $K$ is the number of kernels in the application.
That is, four GPUs are assigned to optimizing each GPU kernel and each optimization loop runs in parallel.
Thus, the smallest run is 28 GPUs for MiniFE and the largest is 216 GPUs for S3D.
We note that each of these frameworks can be run with less or more GPUs (e.g. \rrr can be run on a single GPU); we choose these settings to facilitate fair comparison between the frameworks as each approach can trivially parallelize search across different GPU kernels.

\section{Scaling Evolutionary Optimization Results}\label{sec:evolve-performance-results}
In this section we present our scaling results across our optimizations.
We first show how record-replay turns single thread, sequential evolution from evaluation-bound to inference-bound.
We then show the inference via prefix-cache aware prompting and runtime-aware LLM selection results.

\begin{figure}[ht]
    \centering
    \includegraphics[width=\linewidth]{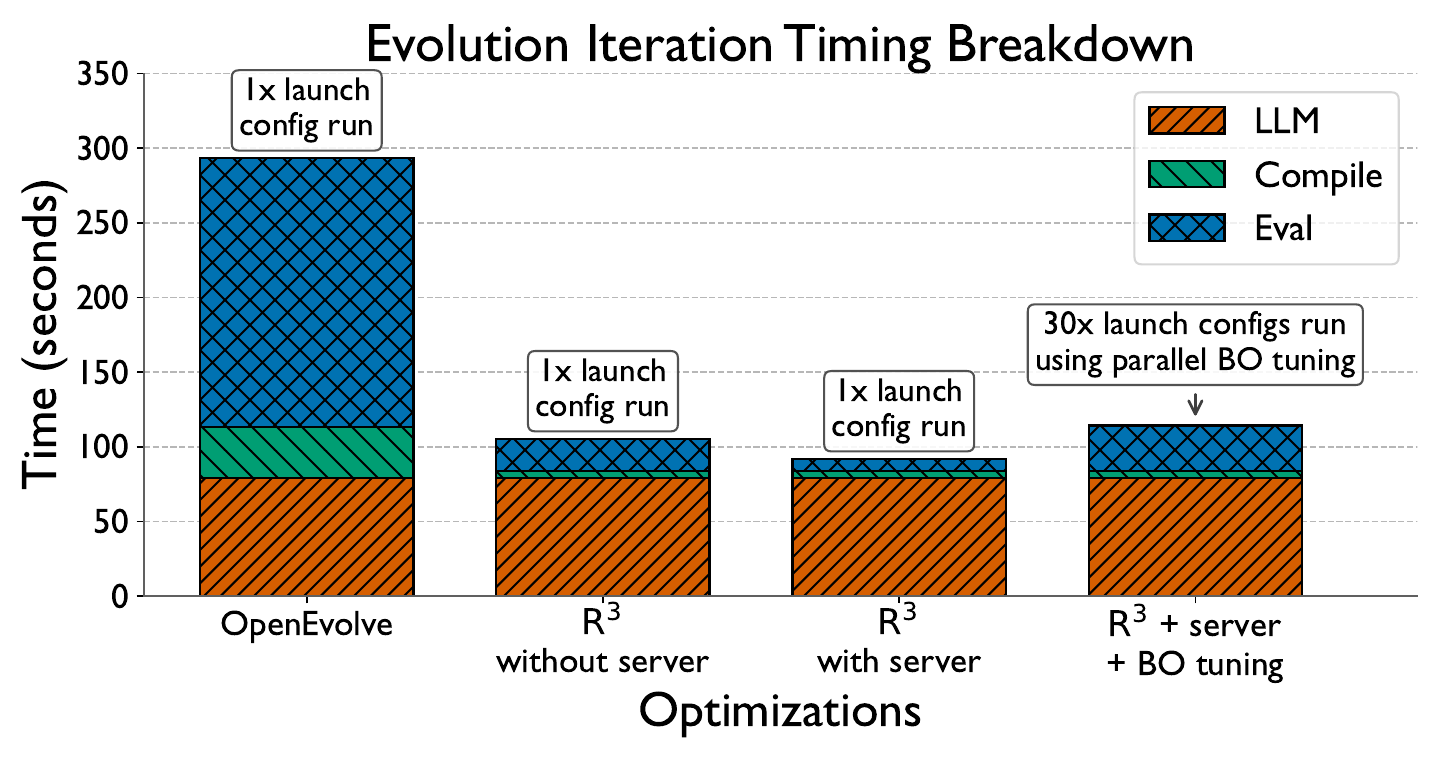}
    \caption{
        The breakdown in times spent in generation, compiling, and evaluation during evolution with each progressive optimization used. Results are shown for 1 sequential evolve process using gpt-oss-120b with high reasoning effort on the QUDA application.
        We see our proposed optimizations have moved evolutionary optimization from evaluation-bound to LLM inference-bound.
    \label{fig:evolve-time-breakdown}}
    \vspace{-1em}
\end{figure}

\Cref{fig:evolve-time-breakdown} shows the breakdown of times for a single evolution iteration when optimizing the LQCD application QUDA~\cite{CLARK20101517} (see~\Cref{sec:case-study}).
We run the evolution five times and average iteration timings from the 10th to 20th iteration across those five evolution runs; this is to control for the fact that evaluation will vary throughout the course of evolution as the kernels are optimized.
Limiting to the 10th to 20th iterations gives us a glimpse across consistent kernel timings, but the trends hold throughout evolution.

We see that replacing standard OpenEvolve evaluation with record-replay drastically reduces compile and evaluation time.
QUDA compile time is reduced by $\approx 86\%$ and eval time by $\approx 88\%$.
Note that the OpenEvolve baseline uses partial compilation; we re-use existing compiled binaries and only recompile source files with changes.
If you wanted to build from scratch, each OpenEvolve evaluation would be significantly slower, as compiling from scratch took $\approx 30$ minutes on our test system.
In \rrr we only need to generate IR for the changed code and can avoid compiler lowering costs and linking costs.
{\bf \emph{Most notably we have transitioned the evolution iteration from evaluation-bound in OpenEvolve to inference-bound in \rrr.}}

\begin{figure}[h]
    \centering
    \includegraphics[width=\linewidth]{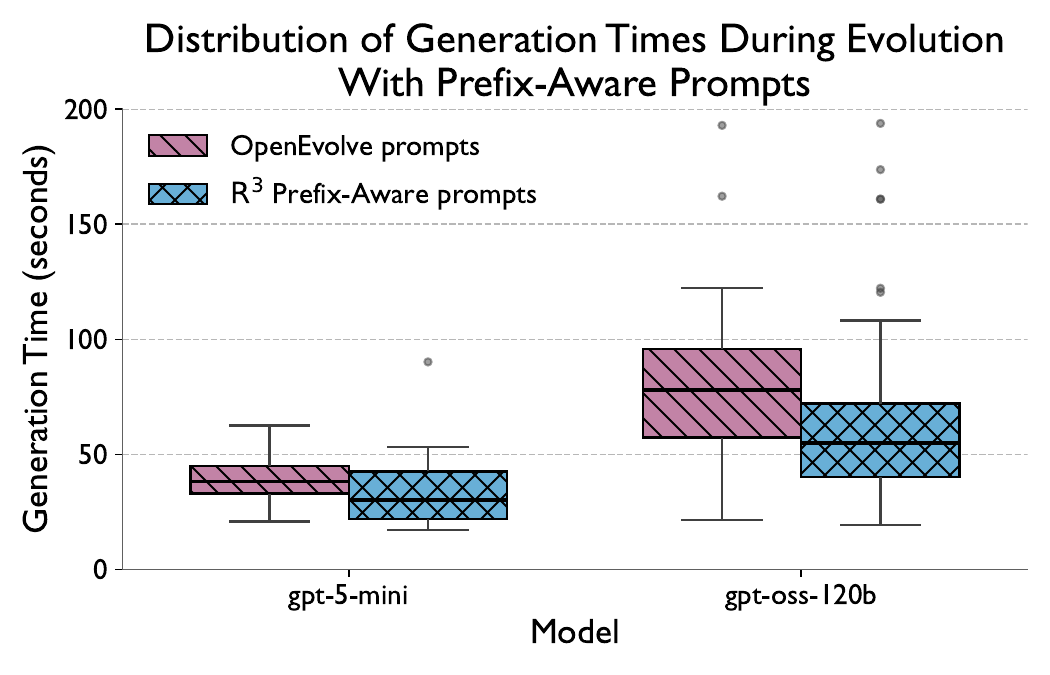}
    \caption{Comparison of time spent generating outputs with and without prefix-aware prompts across a 250 iteration evolution run. We see a reduction in median generation time for both a commercial (gpt-5-mini) and local (gpt-oss-120b) model. Notably, most commercial model providers offer discounts for prefix-cache hits, leading to lower overall costs during evolution. \label{fig:prefix-aware-prompts}}
\end{figure}

Now that we have successfully reduced the evaluation time and are inference-bound, we focus on optimizing time spent in LLM generation.
\Cref{fig:prefix-aware-prompts} shows the distribution of time spent generating outputs across LLMs between OpenEvolve and \rrr in a long 250 iteration run (we exclude gpt-5 from this experiment due to cost).
We see a reduction in generation time for both models when using \rrr prefix-aware prompts. 
Notably, many LLM providers provide discounts for tokens that hit the prefill cache leading to reduced inference costs when using \rrr.
It is possible for inference times to go down due to the reformatting of the prompts in \rrr leading to LLMs generating shorter reasoning traces or final outputs.
To confirm that cache hits are actually what is aiding here, we re-run the experiment, but manually turn off prefix caching in gpt-oss-120b's vLLM server.
The resulting distribution for gpt-oss-120b is the same as OpenEvolve's, confirming that \rrr is actually benefiting from prefix cache optimizations.

\begin{figure}[h]
    \centering
    \includegraphics[width=\linewidth]{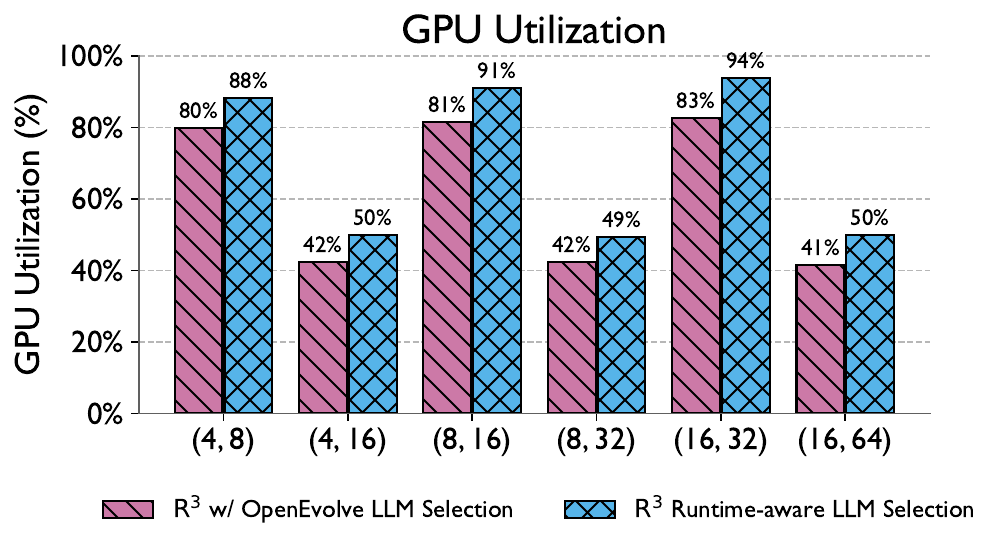}
    \includegraphics[width=\linewidth]{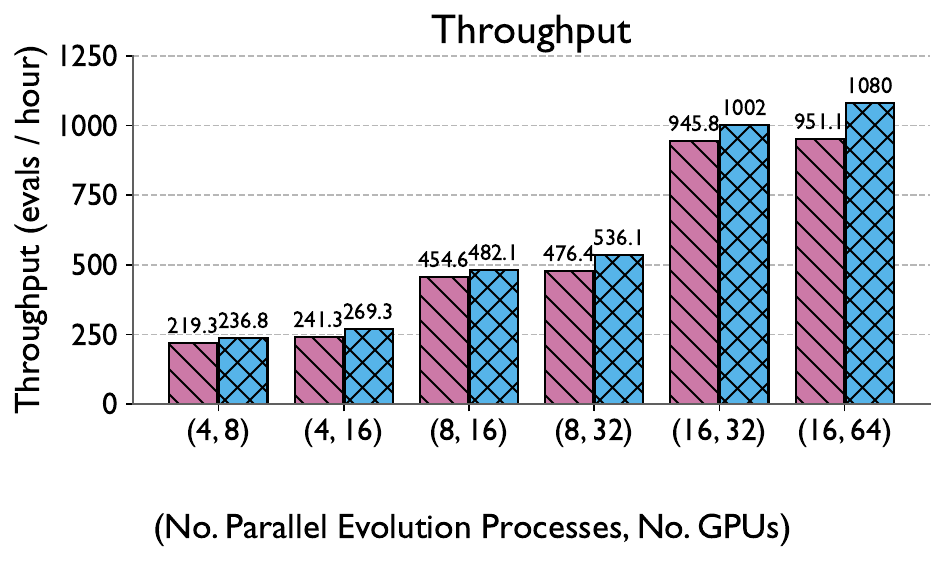}
    \caption{Comparison of OpenEvolve's LLM selection algorithm to \rrr's runtime-aware LLM selection. Whereas increasing the number of parallel evolution processes w.r.t. the number of GPUs can increase utilization and throughput, it comes at the cost of increasing stale MAP-Elites database reads and slowing convergence. \rrr's runtime-aware LLM selection increases utilization and throughput over OpenEvolve for the same number of processes. \label{fig:smart_scaling}}
\end{figure}

\Cref{fig:smart_scaling} further shows how \rrr's runtime-aware LLM selection algorithm improves throughput over OpenEvolve's LLM selection algorithm by up to 13.5\%.
From \Cref{fig:smart_scaling} it is clear to see that increasing the ratio of parallel evolution processes $P$ to number of evaluation GPUs $G$ will increase GPU utilization and throughput.
For example, consider the points (4,16) and (8,16) in~\Cref{fig:smart_scaling}.
With the same number of GPUs for evaluation (16) the throughput and utilization are nearly doubled by increasing $P$ from 4 to 8.
However, as $P$ increases past the number of islands and/or closer to $G$ we get more stale reads from the MAP-Elites database and potentially worse/slower convergence.
If the number of islands is fixed at $4$, as in our experiments, changing $P$ from 4 to 8 will nearly double the number of stale database reads and hurt convergence.
Thus, it is important to improve utilization and throughput without increasing $P$.
We see that \rrr's runtime-aware selection algorithm increases utilization and throughput over OpenEvolve with the same $P$.
In some instances \rrr achieves up to an 11\% increase in GPU utilization or 13.5\% more evaluations per hour.

\section{Results}\label{sec:results}
In this section we present results comparing \rrr with record-replay + BO and OpenEvolve on the four applications.

\begin{figure}[h]
    \centering
    \includegraphics[width=\linewidth]{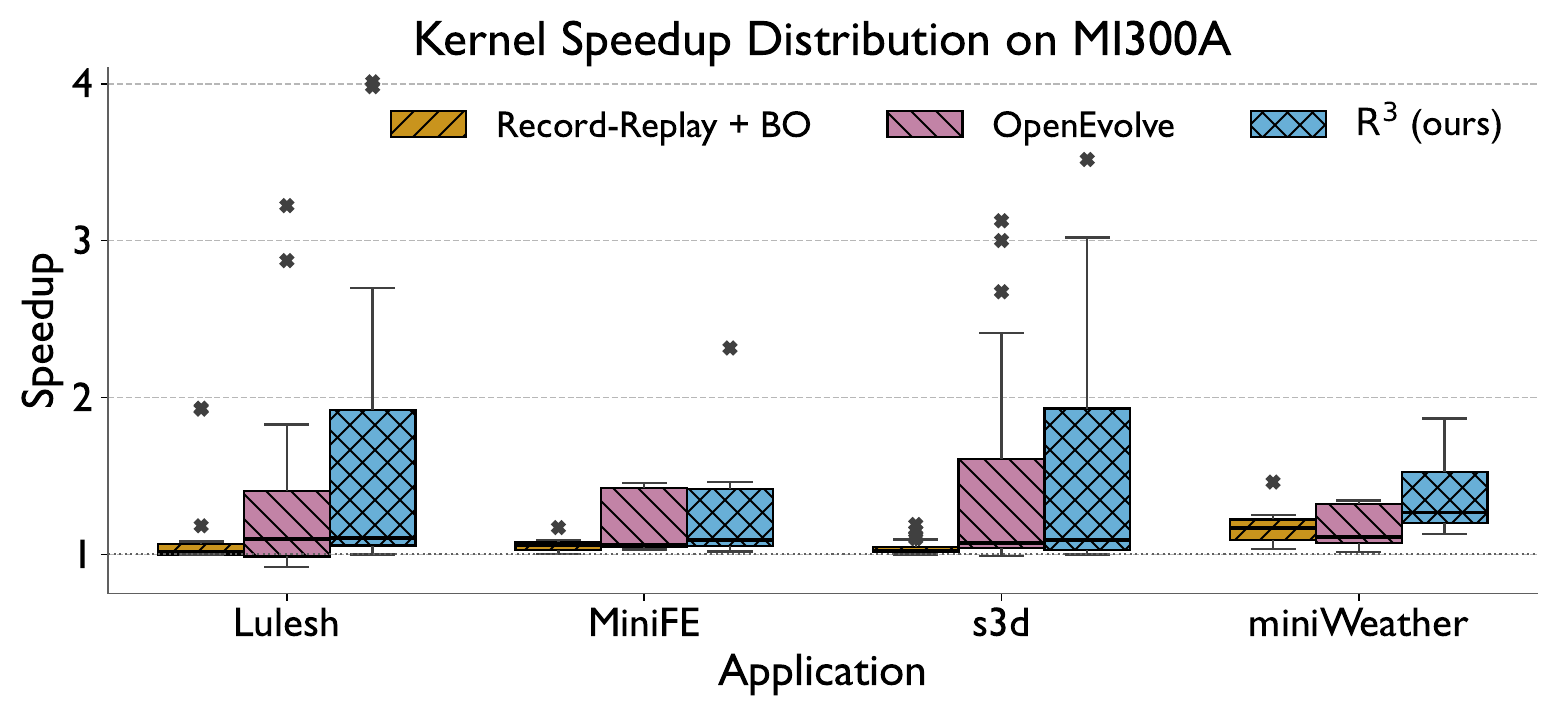}
    \caption{
        Comparison of kernel speedups across the approaches and applications on an MI300A. Times are shown for all kernels in an application (Lulesh 15, MiniFE 7, s3d 54, and miniWeather 7) and dots represent outliers. \rrr yields higher median and max speedups over BO and OpenEvolve.
    \label{fig:kernel-speedups-hip}}
\end{figure}

\begin{figure}[h]
    \centering
    \includegraphics[width=\linewidth]{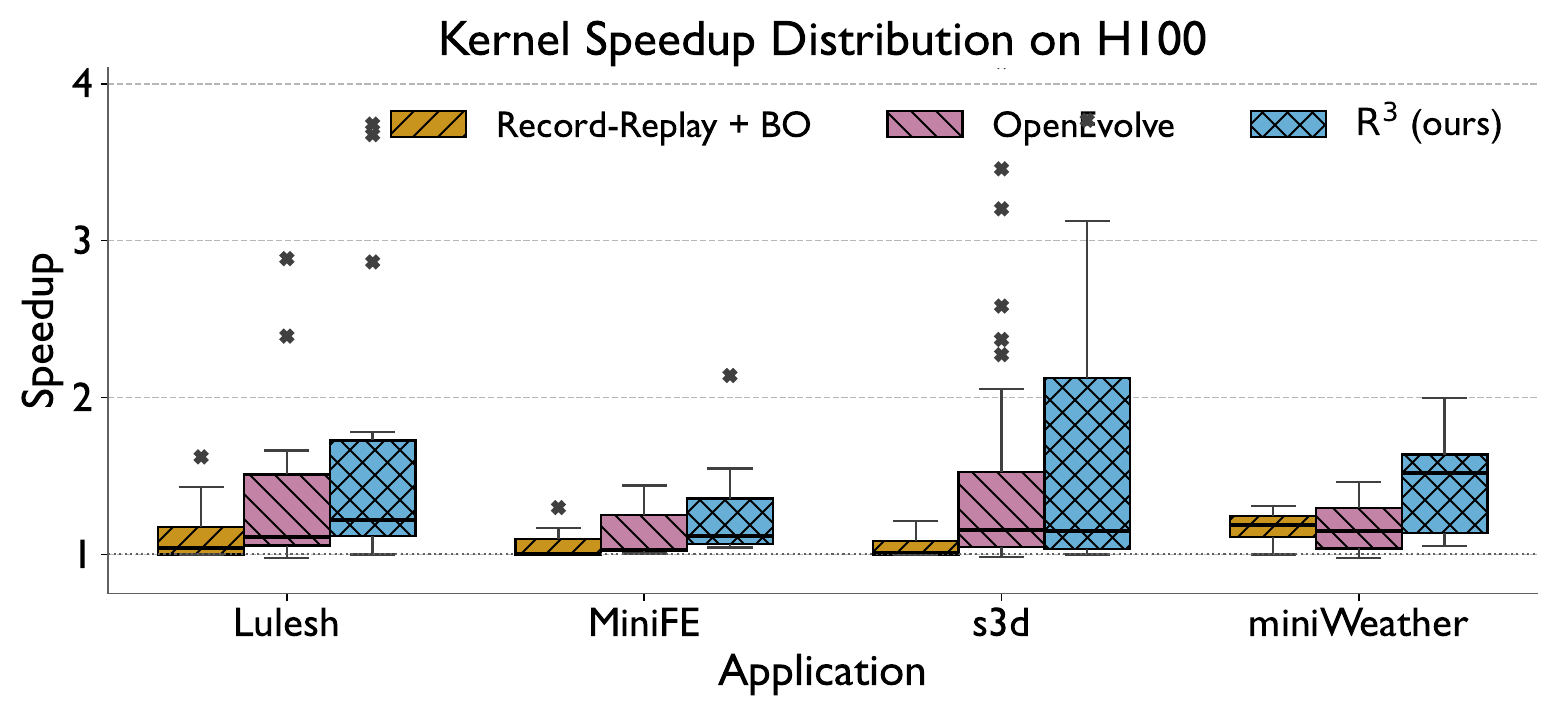}
    \caption{
        Comparison of kernel speedups across the approaches and applications on an H100. Times are shown for all kernels in an application and dots represent outliers. \rrr yields higher median and max speedups over baselines.
    \label{fig:kernel-speedups-cuda}}
\end{figure}

\Cref{fig:kernel-speedups-hip} shows the distribution of kernel speedups on MI300A from each method over the baseline implementation; results are shown as a distribution over the kernels within an application, e.g. Lulesh has 15 kernels.
For all applications \rrr has a higher median and max speedup than OpenEvolve and record-replay + BO.
Both \rrr and OpenEvolve tend to perform stronger than record-replay + BO demonstrating the impact of including source code rewrites in kernel tuning.
The same results are shown on an H100 GPU in~\Cref{fig:kernel-speedups-cuda} where we see the same trends as with AMD: \rrr yields better optimized kernels than the two baseline approaches.

\begin{figure}[h]
    \includegraphics[width=0.49\linewidth]{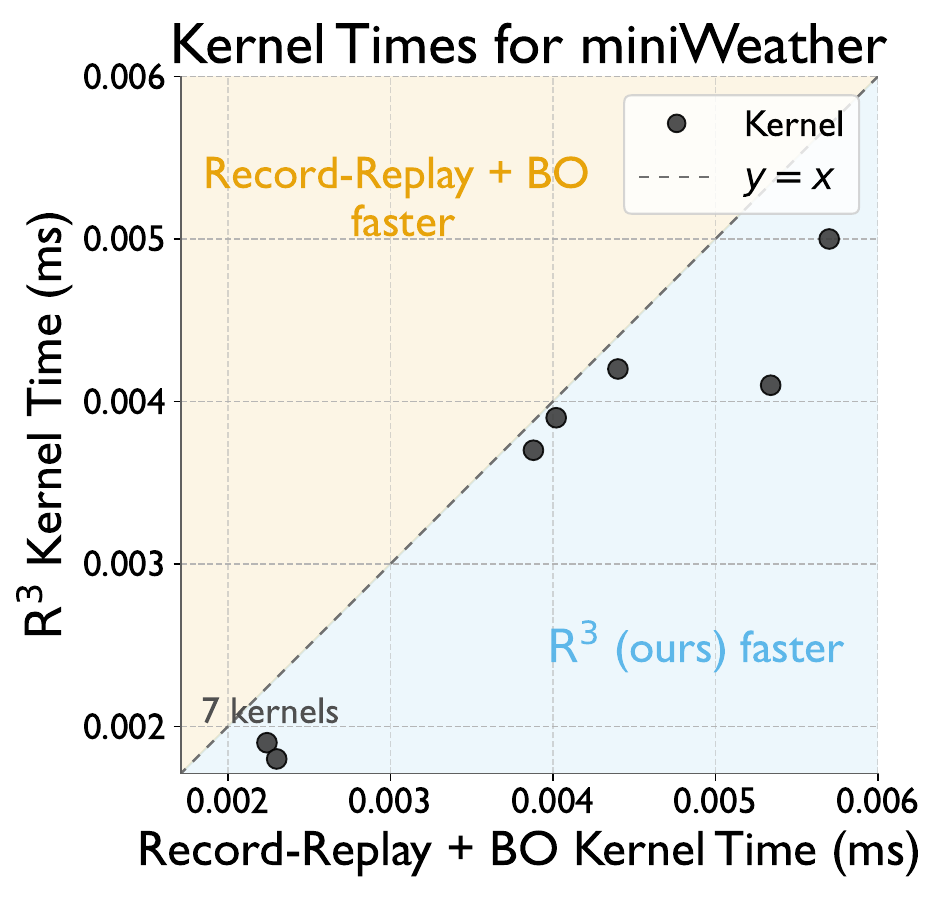}
    \includegraphics[width=0.49\linewidth]{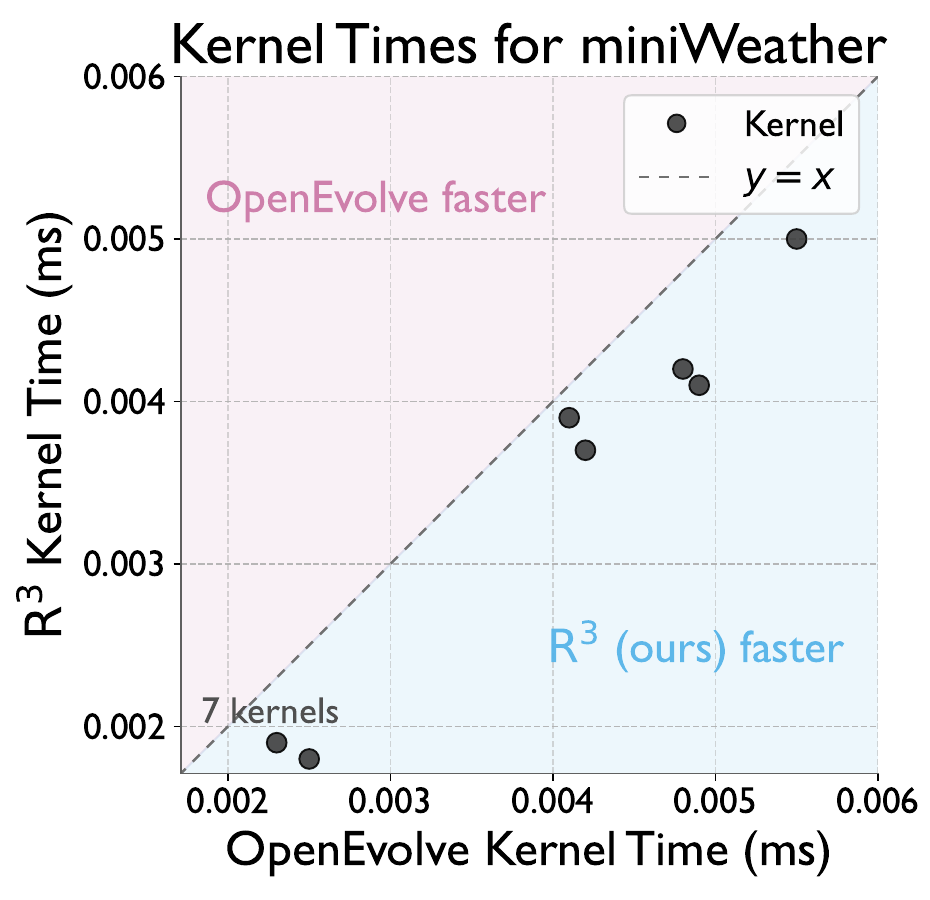}
    \caption{Comparison of absolute times from kernels in the miniWeather application. The y-axis encodes kernel time after \rrr optimization and the x-axis encodes kernel time after optimization from record-replay + BO (left plot) and OpenEvolve (right plot). Points under the line are faster with \rrr. \label{fig:abs-time-compare}}
    \vspace{-1em}
\end{figure}

Not only do the distributions of kernel speedups improve from OpenEvolve to \rrr, but so do the individual kernel times.
\Cref{fig:abs-time-compare} shows the absolute time of the miniWeather kernels as reported by ROCr for all three approaches.
We see that for every kernel \rrr yields a faster optimized kernel than record-replay + BO or OpenEvolve.

\begin{figure}[h]
    \centering
    \includegraphics[width=\linewidth]{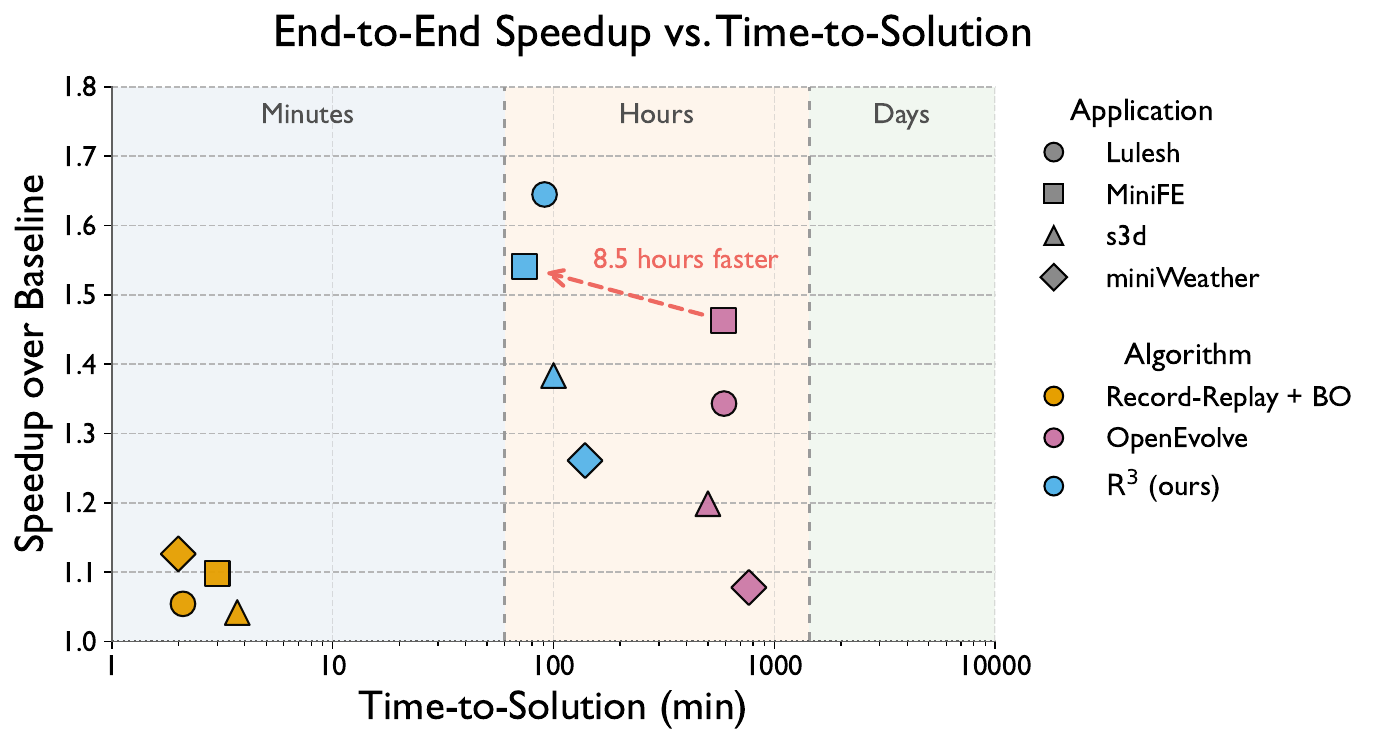}
    \caption{Comparison of final achieved speedup versus the time-to-solution. The vertical axis shows speedup over the original application and the horizontal axis shows time-to-solution in minutes (note log scale). While \rrr is slower than BO with record-replay, it produces consistently better tuning results. Furthermore, \rrr is both faster and yields better speedups than OpenEvolve.
    \label{fig:speedup_vs_time_to_solution}}
\end{figure}

Finally, \Cref{fig:speedup_vs_time_to_solution} shows the speedup of the final application with the optimized kernels from each approach compared with the time-to-solution.
We immediately see two things: (1) \rrr achieves the best overall speedups by combining source level and pass/launch parameter tuning, and (2) \rrr's record-replay engine drastically speeds up evolutionary LLM code optimization.
\rrr is both faster and yields better results than OpenEvolve, while it is slower than record-replay + BO but still yields better speedups.
This is expected as introducing LLMs and another axis of search to tuning will increase time-to-solution.
However, without its optimizations \rrr would take nearly a day to run for a single application.
For MiniFE, \rrr was 8.5 hours faster than OpenEvolve, enabling us to find solutions faster or try more samples in the same allotted time.

\section{Case Study: LQCD}\label{sec:case-study}
We present a case study using \rrr to optimize an expensive kernel in QUDA~\cite{CLARK20101517}, a lattice QCD library, achieving a 28.4\% reduction in compute time on AMD MI300A GPUs.

\subsection{Kernel Description}
QUDA~\cite{CLARK20101517} is a library for lattice quantum chromodynamics (QCD) calculations. 
It employs iterative solvers with adaptive multigrid preconditioning to solve the large sparse linear systems that dominate these simulations.
It is an ideal candidate for stress testing \rrr as it is a full, production HPC application with hundreds of files, tens of thousands of lines of code, long build times, and big workloads.
AlphaEvolve style optimization on this repository would take days making it infeasible for practical use.

Our optimization target is the \texttt{CoarseDslash} kernel, which applies the coarse-level Dirac operator within QUDA's multigrid preconditioner. 
Each thread operates on a single lattice site, computing output spinor values via gauge-covariant nearest-neighbor hopping and local clover contributions. 
During profiling, we find it to be the most expensive kernel in the codebase.
The kernel itself, plus all inline device functions it calls, is large spanning several hundred lines of code.

\subsection{\rrr Optimization Results}
We run \rrr with the same setup as the previous experiments to find optimizations for the \texttt{CoarseDslash} kernel.
In the final optimized kernel, \rrr reorganizes the kernel inner loops so that each input spinor or halo value is loaded once for a given spin/color-column pair and then reused across the output color rows handled by the thread. 
The optimized code also hoists repeatedly used invariants, including references to the gauge and clover fields, halo buffers, source spinors, and row-base indices, out of the hottest loops. 
This reduces redundant field accesses, repeated address arithmetic, and other indexing overhead.
These changes are from LLM source rewrites.

We also observe several optimizations enabled due to the hierarchical nature of \rrr.
Several parts of the kernel are refactored to extract compile-time constants and conditionals, e.g. the dslash/clover selection, warp-fission index calculation, etc.; this enables the record-replay engine's compiler constant specialization optimizations.
\emph{Finding such optimizations would not be possible without a hierarchical search as in \rrr}.

The final kernel speedup on an MI300A is $1.33\times$.
When placed into the whole QUDA application we see a 28.4\% reduction in time spent in compute when run on a representative workload on 64 GPUs (we omit data loading and tear-down time in our profiling).
Due to the time and cost associated with it, we do not run OpenEvolve on the same problem, but based on standalone evaluation times we project it would take just under 2 days (42 hours), while \rrr finished in 108 minutes.

\section{Conclusion}\label{sec:conclusion}

We presented \rrrfull (\rrr), a hierarchical GPU kernel optimization framework that combines LLM-guided evolution for source-level rewrites with replay-based Bayesian optimization for compiler passes and launch configurations. 
By making candidate evaluation fast through record-replay and related systems optimizations, \rrr makes broad search over implementation, compiler, and launch decisions practical. 
Across four scientific applications, \rrr achieves better kernel and application-level performance than record-replay + BO alone, better final results than OpenEvolve, and substantially lower time-to-solution than evolutionary search.

\section*{Acknowledgment}
This work was performed under the auspices
of the U.S.~Department of Energy by Lawrence Livermore National Laboratory
(LLNL) under Contract DE-AC52-07NA27344 (LLNL-CONF-2017736). This work was
supported in part by LLNL LDRD projects 25-ERD-058.
This material is based upon work supported by the U.S. Department of Energy, Office of Science, Office of Advanced Scientific Computing Research, through solicitation DE-FOA-0003264, “Advancements in Artificial Intelligence for Science,” under Award Number DE-SC0025598 and contract DE-AC52-07NA27344.
Computing support for this work came from the Lawrence Livermore National Laboratory
(LLNL) Institutional Computing Grand Challenge program.

\bibliographystyle{IEEEtran}
\bibliography{./bib/zotero-references, ./bib/dinos, bib/extra}

\begin{thebibliography}{10}
\providecommand{\url}[1]{#1}
\csname url@samestyle\endcsname
\providecommand{\newblock}{\relax}
\providecommand{\bibinfo}[2]{#2}
\providecommand{\BIBentrySTDinterwordspacing}{\spaceskip=0pt\relax}
\providecommand{\BIBentryALTinterwordstretchfactor}{4}
\providecommand{\BIBentryALTinterwordspacing}{\spaceskip=\fontdimen2\font plus
\BIBentryALTinterwordstretchfactor\fontdimen3\font minus \fontdimen4\font\relax}
\providecommand{\BIBforeignlanguage}[2]{{%
\expandafter\ifx\csname l@#1\endcsname\relax
\typeout{** WARNING: IEEEtran.bst: No hyphenation pattern has been}%
\typeout{** loaded for the language `#1'. Using the pattern for}%
\typeout{** the default language instead.}%
\else
\language=\csname l@#1\endcsname
\fi
#2}}
\providecommand{\BIBdecl}{\relax}
\BIBdecl

\bibitem{7855909}
\BIBentryALTinterwordspacing
J.~Ansel, S.~Kamil, K.~Veeramachaneni, J.~Ragan-Kelley, J.~Bosboom, U.-M. O'Reilly, and S.~Amarasinghe, ``Opentuner: An extensible framework for program autotuning,'' in \emph{2014 23rd International Conference on Parallel Architecture and Compilation Techniques (PACT)}, 2014, pp. 303--315. [Online]. Available: \url{https://doi.org/10.1145/2628071.2628092}
\BIBentrySTDinterwordspacing

\bibitem{1592880}
\BIBentryALTinterwordspacing
C.~Tapus, I.-H. Chung, and J.~Hollingsworth, ``Active harmony: Towards automated performance tuning,'' in \emph{SC '02: Proceedings of the 2002 ACM/IEEE Conference on Supercomputing}, 2002, pp. 44--44. [Online]. Available: \url{https://doi.org/10.1109/SC.2002.10062}
\BIBentrySTDinterwordspacing

\bibitem{heldens2023kernel}
\BIBentryALTinterwordspacing
S.~Heldens and B.~van Werkhoven, ``Kernel launcher: C++ library for optimal-performance portable cuda applications,'' \emph{arXiv preprint arXiv:2303.12374}, 2023. [Online]. Available: \url{https://doi.org/10.48550/arXiv.2303.12374}
\BIBentrySTDinterwordspacing

\bibitem{7328205}
\BIBentryALTinterwordspacing
C.~Nugteren and V.~Codreanu, ``Cltune: A generic auto-tuner for opencl kernels,'' in \emph{2015 IEEE 9th International Symposium on Embedded Multicore/Many-core Systems-on-Chip (MCSoC)}.\hskip 1em plus 0.5em minus 0.4em\relax Los Alamitos, CA, USA: IEEE Computer Society, sep 2015, pp. 195--202. [Online]. Available: \url{https://doi.org/10.1109/MCSoC.2015.10}
\BIBentrySTDinterwordspacing

\bibitem{10.1007/978-3-319-43659-3_18}
\BIBentryALTinterwordspacing
M.~Popov, C.~Akel, W.~Jalby, and P.~de~Oliveira~Castro, ``Piecewise holistic autotuning of compiler and runtime parameters,'' in \emph{Euro-Par 2016: Parallel Processing}, P.-F. Dutot and D.~Trystram, Eds.\hskip 1em plus 0.5em minus 0.4em\relax Cham: Springer International Publishing, 2016, pp. 238--250. [Online]. Available: \url{https://doi.org/10.1007/978-3-319-43659-3\_18}
\BIBentrySTDinterwordspacing

\bibitem{DBLP:journals/taco/CastroAPPJ15}
\BIBentryALTinterwordspacing
P.~de~Oliveira~Castro, C.~Akel, E.~Petit, M.~Popov, and W.~Jalby, ``{CERE:} llvm-based codelet extractor and replayer for piecewise benchmarking and optimization,'' \emph{{ACM} Trans. Archit. Code Optim.}, vol.~12, no.~1, pp. 6:1--6:24, 2015. [Online]. Available: \url{https://doi.org/10.1145/2724717}
\BIBentrySTDinterwordspacing

\bibitem{10.1145/3572848.3577497}
\BIBentryALTinterwordspacing
Z.~Xie, J.~Liu, J.~Li, and D.~Li, ``Merchandiser: Data placement on heterogeneous memory for task-parallel hpc applications with load-balance awareness,'' in \emph{Proceedings of the 28th ACM SIGPLAN Annual Symposium on Principles and Practice of Parallel Programming}, ser. PPoPP '23.\hskip 1em plus 0.5em minus 0.4em\relax New York, NY, USA: Association for Computing Machinery, 2023, p. 204–217. [Online]. Available: \url{https://doi.org/10.1145/3572848.3577497}
\BIBentrySTDinterwordspacing

\bibitem{parasyris_scalable_20232}
\BIBentryALTinterwordspacing
K.~Parasyris, G.~Georgakoudis, E.~Rangel, I.~Laguna, and J.~Doerfert, ``Scalable {Tuning} of ({OpenMP}) {GPU} {Applications} via {Kernel} {Record} and {Replay},'' in \emph{Proceedings of the {International} {Conference} for {High} {Performance} {Computing}, {Networking}, {Storage} and {Analysis}}, ser. {SC} '23.\hskip 1em plus 0.5em minus 0.4em\relax New York, NY, USA: Association for Computing Machinery, Nov. 2023, pp. 1--14. [Online]. Available: \url{https://doi.org/10.1145/3581784.3607098}
\BIBentrySTDinterwordspacing

\bibitem{liu_survey_20232}
\BIBentryALTinterwordspacing
L.~Liu, T.~Fei, Z.~Zhu, K.~Wu, and Y.~Zhang, ``A {Survey} of {Evolutionary} {Algorithms},'' in \emph{2023 4th {International} {Conference} on {Big} {Data}, {Artificial} {Intelligence} and {Internet} of {Things} {Engineering} ({ICBAIE})}, Aug. 2023, pp. 22--27. [Online]. Available: \url{https://doi.org/10.1109/ICBAIE59714.2023.10281260}
\BIBentrySTDinterwordspacing

\bibitem{romera-paredes_mathematical_20242}
\BIBentryALTinterwordspacing
B.~Romera-Paredes, M.~Barekatain, A.~Novikov, M.~Balog, M.~P. Kumar, E.~Dupont, F.~J.~R. Ruiz, J.~S. Ellenberg, P.~Wang, O.~Fawzi, P.~Kohli, and A.~Fawzi, ``\BIBforeignlanguage{en}{Mathematical discoveries from program search with large language models},'' \emph{\BIBforeignlanguage{en}{Nature}}, vol. 625, no. 7995, pp. 468--475, Jan. 2024. [Online]. Available: \url{https://doi.org/10.1038/s41586-023-06924-6}
\BIBentrySTDinterwordspacing

\bibitem{novikov_alphaevolve_20252}
\BIBentryALTinterwordspacing
A.~Novikov, N.~Vũ, M.~Eisenberger, E.~Dupont, P.-S. Huang, A.~Z. Wagner, S.~Shirobokov, B.~Kozlovskii, F.~J.~R. Ruiz, A.~Mehrabian, M.~P. Kumar, A.~See, S.~Chaudhuri, G.~Holland, A.~Davies, S.~Nowozin, P.~Kohli, and M.~Balog, ``{AlphaEvolve}: {A} coding agent for scientific and algorithmic discovery,'' Jun. 2025. [Online]. Available: \url{https://doi.org/10.48550/arXiv.2506.13131}
\BIBentrySTDinterwordspacing

\bibitem{cheng_barbarians_20252}
\BIBentryALTinterwordspacing
A.~Cheng, S.~Liu, M.~Pan, Z.~Li, B.~Wang, A.~Krentsel, T.~Xia, M.~Cemri, J.~Park, S.~Yang, J.~Chen, L.~Agrawal, A.~Desai, J.~Xing, K.~Sen, M.~Zaharia, and I.~Stoica, ``Barbarians at the {Gate}: {How} {AI} is {Upending} {Systems} {Research},'' Oct. 2025. [Online]. Available: \url{https://doi.org/10.48550/arXiv.2510.06189}
\BIBentrySTDinterwordspacing

\bibitem{openevolve}
\BIBentryALTinterwordspacing
A.~Sharma, ``Openevolve: an open-source evolutionary coding agent,'' 2025. [Online]. Available: \url{https://github.com/algorithmicsuperintelligence/openevolve}
\BIBentrySTDinterwordspacing

\bibitem{mouret_illuminating_20152}
\BIBentryALTinterwordspacing
J.-B. Mouret and J.~Clune, ``Illuminating search spaces by mapping elites,'' Apr. 2015. [Online]. Available: \url{https://doi.org/10.48550/arXiv.1504.04909}
\BIBentrySTDinterwordspacing

\bibitem{van2019kernel}
\BIBentryALTinterwordspacing
B.~van Werkhoven, ``Kernel tuner: A search-optimizing gpu code auto-tuner,'' \emph{Future Generation Computer Systems}, vol.~90, pp. 347--358, 2019. [Online]. Available: \url{https://doi.org/10.1016/j.future.2018.08.004}
\BIBentrySTDinterwordspacing

\bibitem{chen_tvm_20182}
\BIBentryALTinterwordspacing
T.~Chen, T.~Moreau, Z.~Jiang, L.~Zheng, E.~Yan, M.~Cowan, H.~Shen, L.~Wang, Y.~Hu, L.~Ceze, C.~Guestrin, and A.~Krishnamurthy, ``\BIBforeignlanguage{en}{{TVM}: {An} {Automated} {End}-to-{End} {Optimizing} {Compiler} for {Deep} {Learning}},'' Feb. 2018. [Online]. Available: \url{https://doi.org/10.48550/arXiv.1802.04799}
\BIBentrySTDinterwordspacing

\bibitem{zheng_ansor_2020}
\BIBentryALTinterwordspacing
L.~Zheng, C.~Jia, M.~Sun, Z.~Wu, C.~H. Yu, A.~Haj-Ali, Y.~Wang, J.~Yang, D.~Zhuo, K.~Sen, J.~E. Gonzalez, and I.~Stoica, ``Ansor: Generating high-performance tensor programs for deep learning,'' arXiv, 2020, oSDI 2020 (arXiv version). [Online]. Available: \url{https://doi.org/10.48550/arXiv.2006.06762}
\BIBentrySTDinterwordspacing

\bibitem{shao_tensorprob_2022}
\BIBentryALTinterwordspacing
J.~Shao, X.~Zhou, S.~Feng, B.~Hou, R.~Lai, H.~Jin, W.~Lin, M.~Masuda, C.~H. Yu, and T.~Chen, ``Tensor program optimization with probabilistic programs,'' arXiv, 2022, accepted to NeurIPS 2022 (arXiv version). [Online]. Available: \url{https://doi.org/10.48550/arXiv.2205.13603}
\BIBentrySTDinterwordspacing

\bibitem{adams_halide_autoscheduler_2019}
\BIBentryALTinterwordspacing
A.~Adams, K.~Ma, L.~Anderson, R.~Baghdadi, T.-M. Li, M.~Gharbi, B.~Steiner, S.~Johnson, K.~Fatahalian, F.~Durand, and J.~Ragan-Kelley, ``Learning to optimize halide with tree search and random programs,'' \emph{ACM Transactions on Graphics}, vol.~38, no.~4, pp. 121:1--121:12, 2019. [Online]. Available: \url{https://doi.org/10.1145/3306346.3322967}
\BIBentrySTDinterwordspacing

\bibitem{tillet_triton_2019}
\BIBentryALTinterwordspacing
P.~Tillet, H.~T. Kung, and D.~Cox, ``Triton: An intermediate language and compiler for tiled neural network computations,'' in \emph{Proceedings of the 3rd ACM SIGPLAN International Workshop on Machine Learning and Programming Languages (MAPL 2019)}, 2019, pp. 10--19. [Online]. Available: \url{https://doi.org/10.1145/3315508.3329973}
\BIBentrySTDinterwordspacing

\bibitem{10.1145/3124452}
\BIBentryALTinterwordspacing
A.~H. Ashouri, A.~Bignoli, G.~Palermo, C.~Silvano, S.~Kulkarni, and J.~Cavazos, ``Micomp: Mitigating the compiler phase-ordering problem using optimization sub-sequences and machine learning,'' \emph{ACM Trans. Archit. Code Optim.}, vol.~14, no.~3, Sep. 2017. [Online]. Available: \url{https://doi.org/10.1145/3124452}
\BIBentrySTDinterwordspacing

\bibitem{cummins_compilergym_2022}
\BIBentryALTinterwordspacing
C.~Cummins, B.~Wasti, J.~Guo, B.~Cui, J.~Ansel, S.~Gomez, S.~Jain, J.~Liu, O.~Teytaud, B.~Steiner, Y.~Tian, and H.~Leather, ``Compilergym: Robust, performant compiler optimization environments for ai research,'' in \emph{2022 IEEE/ACM International Symposium on Code Generation and Optimization (CGO)}, 2022, pp. 92--105. [Online]. Available: \url{https://doi.org/10.1109/CGO53902.2022.9741258}
\BIBentrySTDinterwordspacing

\bibitem{beckingsale2017apollo}
\BIBentryALTinterwordspacing
D.~Beckingsale, O.~Pearce, I.~Laguna, and T.~Gamblin, ``Apollo: Reusable models for fast, dynamic tuning of input-dependent code,'' in \emph{2017 IEEE International Parallel and Distributed Processing Symposium (IPDPS)}.\hskip 1em plus 0.5em minus 0.4em\relax IEEE, 2017, pp. 307--316. [Online]. Available: \url{https://doi.org/10.1109/IPDPS.2017.38}
\BIBentrySTDinterwordspacing

\bibitem{10.1007/978-3-030-78713-4_24}
\BIBentryALTinterwordspacing
C.~Wood, G.~Georgakoudis, D.~Beckingsale, D.~Poliakoff, A.~Gimenez, K.~Huck, A.~Malony, and T.~Gamblin, ``Artemis: Automatic runtime tuning of parallel execution parameters using machine learning,'' in \emph{High Performance Computing}, B.~L. Chamberlain, A.-L. Varbanescu, H.~Ltaief, and P.~Luszczek, Eds.\hskip 1em plus 0.5em minus 0.4em\relax Cham: Springer International Publishing, 2021, pp. 453--472. [Online]. Available: \url{https://doi.org/10.1007/978-3-030-78713-4\_24}
\BIBentrySTDinterwordspacing

\bibitem{menon_hiperbot_2020}
\BIBentryALTinterwordspacing
H.~Menon, A.~Bhatele, and T.~Gamblin, ``Auto-tuning parameter choices in hpc applications using bayesian optimization,'' in \emph{2020 IEEE International Parallel and Distributed Processing Symposium (IPDPS)}, 2020, pp. 831--840. [Online]. Available: \url{https://doi.org/10.1109/IPDPS47924.2020.00090}
\BIBentrySTDinterwordspacing

\bibitem{10.1145/3437801.3441621}
\BIBentryALTinterwordspacing
Y.~Liu, W.~M. Sid-Lakhdar, O.~Marques, X.~Zhu, C.~Meng, J.~W. Demmel, and X.~S. Li, ``Gptune: Multitask learning for autotuning exascale applications,'' in \emph{Proceedings of the 26th ACM SIGPLAN Symposium on Principles and Practice of Parallel Programming}, ser. PPoPP '21.\hskip 1em plus 0.5em minus 0.4em\relax New York, NY, USA: Association for Computing Machinery, 2021, p. 234–246. [Online]. Available: \url{https://doi.org/10.1145/3437801.3441621}
\BIBentrySTDinterwordspacing

\bibitem{mijakovic_ptf2_2016}
\BIBentryALTinterwordspacing
R.~Mijakovi{\'c}, M.~Firbach, and M.~Gerndt, ``An architecture for flexible auto-tuning: The periscope tuning framework 2.0,'' in \emph{2016 2nd International Conference on Green High Performance Computing (ICGHPC)}, 2016, pp. 1--9. [Online]. Available: \url{https://doi.org/10.1109/ICGHPC.2016.7508066}
\BIBentrySTDinterwordspacing

\bibitem{proteus}
\BIBentryALTinterwordspacing
G.~Georgakoudis, K.~Parasyris, and D.~Beckingsale, ``Proteus: {Portable} {Runtime} {Optimization} of {GPU} {Kernel} {Execution} with {Just}-in-{Time} {Compilation},'' in \emph{Proceedings of the 23rd {ACM}/{IEEE} {International} {Symposium} on {Code} {Generation} and {Optimization}}, ser. {CGO} '25.\hskip 1em plus 0.5em minus 0.4em\relax New York, NY, USA: Association for Computing Machinery, Mar. 2025, pp. 507--522. [Online]. Available: \url{https://doi.org/10.1145/3696443.3708939}
\BIBentrySTDinterwordspacing

\bibitem{nichols2025integratingperformancetoolsmodel}
\BIBentryALTinterwordspacing
D.~Nichols, K.~Parasyris, C.~Jekel, A.~Bhatele, and H.~Menon, ``Integrating performance tools in model reasoning for gpu kernel optimization,'' 2025, https://doi.org/10.48550/arXiv.2510.17158. [Online]. Available: \url{https://arxiv.org/abs/2510.17158}
\BIBentrySTDinterwordspacing

\bibitem{rlpf}
\BIBentryALTinterwordspacing
D.~Nichols, P.~Polasam, H.~Menon, A.~Marathe, T.~Gamblin, and A.~Bhatele, ``{ Performance-Aligned LLMs for Generating Fast HPC Code },'' \emph{IEEE Transactions on Parallel \& Distributed Systems}, no.~01, pp. 1--12, Mar. 5555, https://doi.org/10.1109/TPDS.2026.3675550. [Online]. Available: \url{https://doi.ieeecomputersociety.org/10.1109/TPDS.2026.3675550}
\BIBentrySTDinterwordspacing

\bibitem{kwon_efficient_202322}
\BIBentryALTinterwordspacing
W.~Kwon, Z.~Li, S.~Zhuang, Y.~Sheng, L.~Zheng, C.~H. Yu, J.~E. Gonzalez, H.~Zhang, and I.~Stoica, ``Efficient {Memory} {Management} for {Large} {Language} {Model} {Serving} with {PagedAttention},'' Sep. 2023, arXiv:2309.06180 [cs]. [Online]. Available: \url{https://doi.org/10.48550/arXiv.2309.06180}
\BIBentrySTDinterwordspacing

\bibitem{lulesh}
\BIBentryALTinterwordspacing
I.~Karlin, J.~Keasler, and J.~R. Neely, ``Lulesh 2.0 updates and changes,'' Lawrence Livermore National Laboratory (LLNL), Tech. Rep., 07 2013. [Online]. Available: \url{https://doi.org/10.2172/1090032}
\BIBentrySTDinterwordspacing

\bibitem{lin_assessing_2015}
\BIBentryALTinterwordspacing
P.~T. Lin, M.~A. Heroux, R.~F. Barrett, and A.~B. Williams, ``\BIBforeignlanguage{en}{Assessing a mini-application as a performance proxy for a finite element method engineering application},'' \emph{\BIBforeignlanguage{en}{Concurrency and Computation: Practice and Experience}}, vol.~27, no.~17, pp. 5374--5389, 2015. [Online]. Available: \url{https://doi.org/10.1002/cpe.3587}
\BIBentrySTDinterwordspacing

\bibitem{shoc_s3d}
\BIBentryALTinterwordspacing
A.~Danalis, G.~Marin, C.~McCurdy, J.~S. Meredith, P.~C. Roth, K.~Spafford, V.~Tipparaju, and J.~S. Vetter, ``The scalable heterogeneous computing (shoc) benchmark suite,'' in \emph{Proceedings of the 3rd Workshop on General-Purpose Computation on Graphics Processing Units}, ser. GPGPU-3.\hskip 1em plus 0.5em minus 0.4em\relax New York, NY, USA: Association for Computing Machinery, 2010, p. 63–74. [Online]. Available: \url{https://doi.org/10.1145/1735688.1735702}
\BIBentrySTDinterwordspacing

\bibitem{miniweather}
\BIBentryALTinterwordspacing
M.~R. Norman, ``miniweather,'' [Computer Software], March 2020. [Online]. Available: \url{https://doi.org/10.11578/dc.20201001.88}
\BIBentrySTDinterwordspacing

\bibitem{jin_benchmark_20232}
\BIBentryALTinterwordspacing
Z.~Jin and J.~S. Vetter, ``A {Benchmark} {Suite} for {Improving} {Performance} {Portability} of the {SYCL} {Programming} {Model},'' in \emph{2023 {IEEE} {International} {Symposium} on {Performance} {Analysis} of {Systems} and {Software} ({ISPASS})}, Apr. 2023, pp. 325--327. [Online]. Available: \url{https://doi.org/10.1109/ISPASS57527.2023.00041}
\BIBentrySTDinterwordspacing

\bibitem{bergstra_algorithms_20112}
\BIBentryALTinterwordspacing
J.~Bergstra, R.~Bardenet, Y.~Bengio, and B.~Kégl, ``Algorithms for hyper-parameter optimization,'' in \emph{Proceedings of the 25th {International} {Conference} on {Neural} {Information} {Processing} {Systems}}, ser. {NIPS}'11.\hskip 1em plus 0.5em minus 0.4em\relax Red Hook, NY, USA: Curran Associates Inc., Dec. 2011, pp. 2546--2554, https://dl.acm.org/doi/10.5555/2986459.2986743. [Online]. Available: \url{https://papers.nips.cc/paper/4443-algorithms-for-hyper-parameter-optimization}
\BIBentrySTDinterwordspacing

\bibitem{openai_gpt-oss-120b_20252}
\BIBentryALTinterwordspacing
OpenAI, S.~Agarwal, L.~Ahmad, J.~Ai, S.~Altman, A.~Applebaum, E.~Arbus, R.~K. Arora, Y.~Bai, B.~Baker, H.~Bao, B.~Barak, A.~Bennett, T.~Bertao, N.~Brett, E.~Brevdo, G.~Brockman, S.~Bubeck, C.~Chang, K.~Chen, M.~Chen, E.~Cheung, A.~Clark, D.~Cook, M.~Dukhan, C.~Dvorak, K.~Fives, V.~Fomenko, T.~Garipov, K.~Georgiev, M.~Glaese, T.~Gogineni, A.~Goucher, L.~Gross, K.~G. Guzman, J.~Hallman, J.~Hehir, J.~Heidecke, A.~Helyar, H.~Hu, R.~Huet, J.~Huh, S.~Jain, Z.~Johnson, C.~Koch, I.~Kofman, D.~Kundel, J.~Kwon, V.~Kyrylov, E.~Y. Le, G.~Leclerc, J.~P. Lennon, S.~Lessans, M.~Lezcano-Casado, Y.~Li, Z.~Li, J.~Lin, J.~Liss, Lily, Liu, J.~Liu, K.~Lu, C.~Lu, Z.~Martinovic, L.~McCallum, J.~McGrath, S.~McKinney, A.~McLaughlin, S.~Mei, S.~Mostovoy, T.~Mu, G.~Myles, A.~Neitz, A.~Nichol, J.~Pachocki, A.~Paino, D.~Palmie, A.~Pantuliano, G.~Parascandolo, J.~Park, L.~Pathak, C.~Paz, L.~Peran, D.~Pimenov, M.~Pokrass, E.~Proehl, H.~Qiu, G.~Raila, F.~Raso, H.~Ren, K.~Richardson, D.~Robinson, B.~Rotsted, H.~Salman, S.~Sanjeev,
  M.~Schwarzer, D.~Sculley, H.~Sikchi, K.~Simon, K.~Singhal, Y.~Song, D.~Stuckey, Z.~Sun, P.~Tillet, S.~Toizer, F.~Tsimpourlas, N.~Vyas, E.~Wallace, X.~Wang, M.~Wang, O.~Watkins, K.~Weil, A.~Wendling, K.~Whinnery, C.~Whitney, H.~Wong, L.~Yang, Y.~Yang, M.~Yasunaga, K.~Ying, W.~Zaremba, W.~Zhan, C.~Zhang, B.~Zhang, E.~Zhang, and S.~Zhao, ``gpt-oss-120b \& gpt-oss-20b {Model} {Card},'' Aug. 2025, arXiv:2508.10925 [cs]. [Online]. Available: \url{https://doi.org/10.48550/arXiv.2508.10925}
\BIBentrySTDinterwordspacing

\bibitem{singh_openai_20252}
\BIBentryALTinterwordspacing
A.~Singh, A.~Fry, A.~Perelman, A.~Tart, A.~Ganesh, A.~El-Kishky, A.~McLaughlin, A.~Low, A.~J. Ostrow, A.~Ananthram, A.~Nathan, A.~Luo, A.~Helyar, A.~Madry, A.~Efremov, A.~Spyra, A.~Baker-Whitcomb, A.~Beutel, A.~Karpenko, A.~Makelov, A.~Neitz, A.~Wei, A.~Barr, A.~Kirchmeyer, A.~Ivanov, A.~Christakis, A.~Gillespie, A.~Tam, A.~Bennett, A.~Wan, A.~Huang, A.~M. Sandjideh, A.~Yang, A.~Kumar, A.~Saraiva, A.~Vallone, A.~Gheorghe, A.~G. Garcia, A.~Braunstein, A.~Liu, A.~Schmidt, A.~Mereskin, A.~Mishchenko, A.~Applebaum, A.~Rogerson, A.~Rajan, A.~Wei, A.~Kotha, A.~Srivastava, A.~Agrawal, A.~Vijayvergiya, A.~Tyra, A.~Nair, A.~Nayak, B.~Eggers, B.~Ji, B.~Hoover, B.~Chen, B.~Chen, B.~Barak, B.~Minaiev, B.~Hao, B.~Baker, B.~Lightcap, B.~McKinzie, B.~Wang, B.~Quinn, B.~Fioca, B.~Hsu, B.~Yang, B.~Yu, B.~Zhang, B.~Brenner, C.~R. Zetino, C.~Raymond, C.~Lugaresi, C.~Paz, C.~Hudson, C.~Whitney, C.~Li, C.~Chen, C.~Cole, C.~Voss, C.~Ding, C.~Shen, C.~Huang, C.~Colby, C.~Hallacy, C.~Koch, C.~Lu, C.~Kaplan, C.~Kim, C.~J.
  Minott-Henriques, C.~Frey, C.~Yu, C.~Czarnecki, C.~Reid, C.~Wei, C.~Decareaux, C.~Scheau, C.~Zhang, C.~Forbes, D.~Tang, D.~Goldberg, D.~Roberts, D.~Palmie, D.~Kappler, D.~Levine, D.~Wright, D.~Leo, D.~Lin, D.~Robinson, D.~Grabb, D.~Chen, D.~Lim, D.~Salama, D.~Bhattacharjee, D.~Tsipras, D.~Li, D.~Yu, D.~J. Strouse, D.~Williams, D.~Hunn, E.~Bayes, E.~Arbus, E.~Akyurek, E.~Y. Le, E.~Widmann, E.~Yani, E.~Proehl, E.~Sert, E.~Cheung, E.~Schwartz, E.~Han, E.~Jiang, E.~Mitchell, E.~Sigler, E.~Wallace, E.~Ritter, E.~Kavanaugh, E.~Mays, E.~Nikishin, F.~Li, F.~P. Such, F.~d. A.~B. Peres, F.~Raso, F.~Bekerman, F.~Tsimpourlas, F.~Chantzis, F.~Song, F.~Zhang, G.~Raila, G.~McGrath, G.~Briggs, G.~Yang, G.~Parascandolo, G.~Chabot, G.~Kim, G.~Zhao, G.~Valiant, G.~Leclerc, H.~Salman, H.~Wang, H.~Sheng, H.~Jiang, H.~Wang, H.~Jin, H.~Sikchi, H.~Schmidt, H.~Aspegren, H.~Chen, H.~Qiu, H.~Lightman, I.~Covert, I.~Kivlichan, I.~Silber, I.~Sohl, I.~Hammoud, I.~Clavera, I.~Lan, I.~Akkaya, I.~Kostrikov, I.~Kofman, I.~Etinger,
  I.~Singal, J.~Hehir, J.~Huh, J.~Pan, J.~Wilczynski, J.~Pachocki, J.~Lee, J.~Quinn, J.~Kiros, J.~Kalra, J.~Samaroo, J.~Wang, J.~Wolfe, J.~Chen, J.~Wang, J.~Harb, J.~Han, J.~Wang, J.~Zhao, J.~Chen, J.~Yang, J.~Tworek, J.~Chand, J.~Landon, J.~Liang, J.~Lin, J.~Liu, J.~Wang, J.~Tang, J.~Yin, J.~Jang, J.~Morris, J.~Flynn, J.~Ferstad, J.~Heidecke, J.~Fishbein, J.~Hallman, J.~Grant, J.~Chien, J.~Gordon, J.~Park, J.~Liss, J.~Kraaijeveld, J.~Guay, J.~Mo, J.~Lawson, J.~McGrath, J.~Vendrow, J.~Jiao, J.~Lee, J.~Steele, J.~Wang, J.~Mao, K.~Chen, K.~Hayashi, K.~Xiao, K.~Salahi, K.~Wu, K.~Sekhri, K.~Sharma, K.~Singhal, K.~Li, K.~Nguyen, K.~Gu-Lemberg, K.~King, K.~Liu, K.~Stone, K.~Yu, K.~Ying, K.~Georgiev, K.~Lim, K.~Tirumala, K.~Miller, L.~Ahmad, L.~Lv, L.~Clare, L.~Fauconnet, L.~Itow, L.~Yang, L.~Romaniuk, L.~Anise, L.~Byron, L.~Pathak, L.~Maksin, L.~Lo, L.~Ho, L.~Jing, L.~Wu, L.~Xiong, L.~Mamitsuka, L.~Yang, L.~McCallum, L.~Held, L.~Bourgeois, L.~Engstrom, L.~Kuhn, L.~Feuvrier, L.~Zhang, L.~Switzer, L.~Kondraciuk,
  L.~Kaiser, M.~Joglekar, M.~Singh, M.~Shah, M.~Stratta, M.~Williams, M.~Chen, M.~Sun, M.~Cayton, M.~Li, M.~Zhang, M.~Aljubeh, M.~Nichols, M.~Haines, M.~Schwarzer, M.~Gupta, M.~Shah, M.~Huang, M.~Dong, M.~Wang, M.~Glaese, M.~Carroll, M.~Lampe, M.~Malek, M.~Sharman, M.~Zhang, M.~Wang, M.~Pokrass, M.~Florian, M.~Pavlov, M.~Wang, M.~Chen, M.~Wang, M.~Feng, M.~Bavarian, M.~Lin, M.~Abdool, M.~Rohaninejad, N.~Soto, N.~Staudacher, N.~LaFontaine, N.~Marwell, N.~Liu, N.~Preston, N.~Turley, N.~Ansman, N.~Blades, N.~Pancha, N.~Mikhaylin, N.~Felix, N.~Handa, N.~Rai, N.~Keskar, N.~Brown, O.~Nachum, O.~Boiko, O.~Murk, O.~Watkins, O.~Gleeson, P.~Mishkin, P.~Lesiewicz, P.~Baltescu, P.~Belov, P.~Zhokhov, P.~Pronin, P.~Guo, P.~Thacker, Q.~Liu, Q.~Yuan, Q.~Liu, R.~Dias, R.~Puckett, R.~Arora, R.~T. Mullapudi, R.~Gaon, R.~Miyara, R.~Song, R.~Aggarwal, R.~J. Marsan, R.~Yemiru, R.~Xiong, R.~Kshirsagar, R.~Nuttall, R.~Tsiupa, R.~Eldan, R.~Wang, R.~James, R.~Ziv, R.~Shu, R.~Nigmatullin, S.~Jain, S.~Talaie, S.~Altman, S.~Arnesen,
  S.~Toizer, S.~Toyer, S.~Miserendino, S.~Agarwal, S.~Yoo, S.~Heon, S.~Ethersmith, S.~Grove, S.~Taylor, S.~Bubeck, S.~Banesiu, S.~Amdo, S.~Zhao, S.~Wu, S.~Santurkar, S.~Zhao, S.~R. Chaudhuri, S.~Krishnaswamy, Shuaiqi, Xia, S.~Cheng, S.~Anadkat, S.~P. Fishman, S.~Tobin, S.~Fu, S.~Jain, S.~Mei, S.~Egoian, S.~Kim, S.~Golden, S.~Q. Mah, S.~Lin, S.~Imm, S.~Sharpe, S.~Yadlowsky, S.~Choudhry, S.~Eum, S.~Sanjeev, T.~Khan, T.~Stramer, T.~Wang, T.~Xin, T.~Gogineni, T.~Christianson, T.~Sanders, T.~Patwardhan, T.~Degry, T.~Shadwell, T.~Fu, T.~Gao, T.~Garipov, T.~Sriskandarajah, T.~Sherbakov, T.~Kaftan, T.~Hiratsuka, T.~Wang, T.~Song, T.~Zhao, T.~Peterson, V.~Kharitonov, V.~Chernova, V.~Kosaraju, V.~Kuo, V.~Pong, V.~Verma, V.~Petrov, W.~Jiang, W.~Zhang, W.~Zhou, W.~Xie, W.~Zhan, W.~McCabe, W.~DePue, W.~Ellsworth, W.~Bain, W.~Thompson, X.~Chen, X.~Qi, X.~Xiang, X.~Shi, Y.~Dubois, Y.~Yu, Y.~Khakbaz, Y.~Wu, Y.~Qian, Y.~T. Lee, Y.~Chen, Y.~Zhang, Y.~Xiong, Y.~Tian, Y.~Cha, Y.~Bai, Y.~Yang, Y.~Yuan, Y.~Li, Y.~Zhang, Y.~Yang,
  Y.~Jin, Y.~Jiang, Y.~Wang, Y.~Wang, Y.~Liu, Z.~Stubenvoll, Z.~Dou, Z.~Wu, and Z.~Wang, ``{OpenAI} {GPT}-5 {System} {Card},'' Dec. 2025, arXiv:2601.03267 [cs]. [Online]. Available: \url{https://doi.org/10.48550/arXiv.2601.03267}
\BIBentrySTDinterwordspacing

\bibitem{CLARK20101517}
\BIBentryALTinterwordspacing
M.~Clark, R.~Babich, K.~Barros, R.~Brower, and C.~Rebbi, ``Solving lattice qcd systems of equations using mixed precision solvers on gpus,'' \emph{Computer Physics Communications}, vol. 181, no.~9, pp. 1517--1528, 2010. [Online]. Available: \url{https://doi.org/10.1016/j.cpc.2010.05.002}
\BIBentrySTDinterwordspacing

\end{thebibliography}

\end{document}